# Some Worlds of Quantum Theory

Jeremy Butterfield

All Souls College, Oxford OX1 4AL: email jb56@cus.cam.ac.uk

Abstract: This paper assesses the Everettian approach to the measurement problem, especially the version of that approach advocated by Simon Saunders and David Wallace. I emphasise conceptual, indeed metaphysical, aspects rather than technical ones; but I include an introductory exposition of decoherence. In particular, I discuss whether---as these authors maintain---it is acceptable to have no precise definition of 'branch' (in the Everettian kind of sense). (A version of this paper will appear in a CTNS/Vatican Observatory volume on Quantum Theory and Divine Action, ed. Robert Russell et al.)

## 1. Introduction

In this paper, I shall sketch some of the issues that arise in the interpretation of quantum theory, from the perspective of a philosopher of physics. I will emphasise the measurement problem: for it is widely agreed to be the theory's main interpretative problem (and to equally confront advanced quantum theories, like quantum field theory). My over-arching aim will be to emphasise how hard it is to interpret quantum theory satisfactorily: or in other (more positive!) words, how strange are the various ontologies that the endeavour of interpreting quantum theory suggests.

   After some introductory remarks (Sections 1 and 2), I present the measurement problem (Section 3). I emphasise the role of decoherence (Section 4). Then I distinguish four broad strategies for solving the measurement problem, and briefly describe three of them (Section 5). In the last three Sections (6, 7 and 8), I discuss in more detail the fourth strategy, viz. that of Everett (1957). This strategy encompasses several different interpretations, but I will focus on those of Saunders (1995, 1996, 1996a, 1998) and Wallace (2001, 2001a). (The interpretations of Deutsch, Gell-Mann & Hartle, Vaidman and Zurek are similar---at least as I read them! But I will not enter into details of their views.) My rationale for this choice is that the Everettian strategy has various merits, and illustrates well how radical we might have to be in interpreting quantum theory. Besides, the positions of Saunders and Wallace seem to me among the most attractive versions of the Everettian strategy.

   For the most part, I will cast the discussion (of all the strategies) in terms of elementary quantum theory, neglecting relativity (whether special or general) and quantum field theory; indeed, the discussion will be philosophical, with almost no use even of elementary quantum theory's technicalities. This restriction will not, I think, be misleading, for three reasons. First, most of the discussion---both about the measurement problem and about the strategies for solving it---also applies, once we take into account relativity and quantum field theory.

The second reason relates to my over-arching aim of emphasising the difficulties of interpretation. For these difficulties are in general aggravated once we consider relativity and quantum field theory. I shall occasionally give examples of these aggravations, but we can already note two brief points. (a) The measurement problem is liable to be *more* severe, once we accept, with relativity, that there is no absolute simultaneity, since one has to ask relative to which notion of simultaneity does the 'wave-packet collapse'. (b) By and large, the four broad interpretative strategies face new difficulties once we turn to relativity, e.g. if wave-packet collapse is a physical process, one has to define an appropriately relativistic dynamics.

The third reason relates to my focussing from Section 6 onwards on the fourth strategy, the Everettian one. There is general agreement that this strategy is more easily adapted to relativity and quantum field theory than are the other three. For it does not propose to solve the measurement problem by changes to the formalism of quantum theory: while some of the other strategies' proposed changes are hard to adapt to relativity and quantum field theory, however satisfactory they seem for elementary quantum theory. The Everettian strategy proposes instead to solve the measurement problem by distinctively interpretative, indeed philosophical, claims: claims which, by and large, one can reasonably expect to carry over smoothly to relativity (and quantum field theory)---if one accepts that they work for the non-relativistic case.

I should add that (partly due to lack of space) the paper will have other limitations. First, I address only some aspects of the Everettian strategy. In part this is because I have discussed other versions than those of Saunders and Wallace elsewhere (1995, 1996). But even as regards their versions, I will say only a little about identity through time (in Section 8), since I largely agree with what they say about it. Besides, I will say next to nothing about two other aspects: (i) the interpretation of probability, since again I largely agree with Saunders' treatment; (ii) the consistent histories approach, in terms of which Saunders (though not Wallace) casts his views---for I want to advertise Saunders' and Wallace's main ideas, and they can be conveyed without using the technicalities of consistent histories. (Clarke (this volume) discusses consistent histories.)

The paper will have three other more general limitations. (1): I will say nothing about what connections the interpretation of quantum theory might have with theological issues. (2): I will say nothing about issues arising from the curious non-local phenomena that are described by quantum theory; (phenomena which occur just as much in the context of relativistic quantum theory, as in non-relativistic quantum theory). (3): I will say next to nothing about three out of my four broad strategies for solving the measurement problem.

My reasons for these limitations are straightforward, but connected. As to (1), I am not competent. As to (2), Redhead (this volume) will address these issues. As to (3), I think that by and large, the strategies that I neglect involve less strange ontologies than the fourth strategy on which I concentrate. To that extent they are more credible; but as we shall see, they face other difficulties. In any case, Cushing (this volume) will discuss in detail what is perhaps the best-developed example of these other strategies; viz. the pilot-wave interpretation of quantum theory (and of quantum field theory).

There are also of course connections between the topics of (1) to (3). Most accounts of quantum theory, especially in the popular science literature, say that it is indeterministic (the 'collapse of the wave-packet'); so a theologian naturally asks whether this might afford some scope for divine action in the world. But beware: this much-touted indeterminism is very questionable. Only some of the four strategies accept it. In particular, the pilot-wave interpretation is a perfectly tenable, and utterly deterministic, interpretation; and in a similar vein, the Everettian strategy postulates a fundamental determinism, overlaid by an apparent indeterminism. (Besides, some of the indeterministic interpretations give such detailed and thoroughly physical models of the collapse of the wave-packet, that a theologian might well think twice about finding in them much scope for divine action.)

I should also warn against what one might call the opposite error to that just mentioned: namely, assuming that quantum theory is *merely* indeterministic---and thus that the theology of divine action can proceed, as regards its consideration of quantum theory, with only that notion. As I hinted above, I think this assumption is false: setting aside the pilot-wave interpretation, the other tenable interpretations of quantum theory each involve a stranger ontology than just the notion of indeterminism conveys. To put the point more pithily, in terms of reality rather than interpretations of a theory: quantum theory intimates that reality is stranger than just indeterminism suggests.

Admittedly, it is frustratingly hard to say what this reality might be like, let alone agree on such a description: witness the effort that goes into articulating the various interpretations, and the continuing controversy about which is right. But I hope that the paper will give some indication of what strange metaphysical possibilities are 'on the cards'.

## 2. The Successes of Quantum Physics---and its Open Problems

Before embarking on the measurement problem, I propose to set the stage in two ways. First, I will emphasise how extraordinarily successful quantum theory has proved to be in domains of application far beyond its original one. The moral of this success is that here is a theory we must 'take very seriously' as describing the world. This moral leads to my second topic: some cautionary remarks about quantum theory's remaining problems, and about scientific realism and reductionism.

Examples of quantum theory's success are legion: to take just one, quantum theory is needed to describe the laser in a CD player. More generally, although quantum theory was devised for systems of atomic dimensions ($10^{-8}$ cm.), it turned out to be good both for scales much smaller (cf. the nuclear radius of ca. $10^{-12}$ cm.) and vastly larger (cf. superconductivity and superfluidity, involving scales up to $10^{-1}$ cm.). Indeed, much of the history of twentieth century physics is the story of the consolidation of the quantum revolution: the story of quantum theory's basic postulates being successfully applied ever more widely.

The point applies equally well when we look beyond physics on earth in the present era. Quantum theory has been extraordinarily successful in application to astronomy: the obvious example is the use of nuclear physics to develop a very accurate and detailed theory of stellar structure and evolution, including the creation of nuclei in stars. And similarly in cosmology: we now know a lot about primordial nucleosynthesis.

And yet: complacency, let alone triumphalism, is not in order! Not only is physics full of unfinished business: that is always true in human enquiry. There are also two more specific points, which I should note.

First, there is of course no valid inference from a theory's empirical success, even so wide and detailed a success as quantum theory's, to its literal truth. The most that might be inferred is that it is rational to believe the theory is true, or approximately true; and even that inference requires a strong form of that contentious doctrine, scientific realism. Secondly, apart from these philosophical generalities, physics itself contains hints that quantum theory is not true. That is, there are 'clouds on the horizon' that may prove as great a threat to the continued success of quantum theory (and more generally, of twentieth century physics), as were the anomalies confronting classical physics at the end of the nineteenth century. Of course, people differ in what problems they find worrisome. As a philosopher, I of course find the various mysteries of interpreting quantum theory worrisome, as discussed in later Sections. But I should mention that the wider physics community also recognizes at least two other 'clouds': the need to somehow reconcile quantum theory and general relativity; and the cluster of problems about such issues as renormalizability and the rigorous existence of quantum field theories---i.e. problems about the relation of quantum theory to *special* relativity. (Butterfield (2001) briefly discusses these clouds from a philosophical perspective.)

Fortunately, these points do not undermine this paper's project, of articulating and debating interpretations of quantum theory. For this project does not require one to ignore the clouds that suggest quantum theory has limitations; (let alone to presuppose scientific realism). On the contrary, one should take account of them, if one can: otherwise interpretative debate is likely to become both isolated from the data of physics (to its detriment), and silent as regards heuristic ideas for physics (to the detriment of physics).

However, for the reasons given in Section 1, this paper will concentrate on elementary quantum theory.

Finally, by way of stage-setting, I need to make two remarks about the relation between theories. (1): The first is a cautionary remark about reductionism; which is, I think, important, albeit obvious. I believe there is some strong sense in which particle physics, as nowadays described using quantum field theory, underlies atomic physics; and that atomic physics, in a broadly similar way, underlies chemistry; which in a broadly similar way, underlies biochemistry; and so on.

But these inter-theoretic relations vary from one example to another, and are in any single example often subtle and complicated, e.g. because they involve various kinds of limit: cf. Berry (this volume, 1991) for further discussion, especially of the importance of singular limits. So such examples will not provide neat cases of the traditional philosophical conception of theory-reduction as logical definitional extension; (i.e. as deducing one theory from the other, as supplemented with judiciously chosen definitions of terms). Furthermore, to understand any single example we may well have to take a stand on issues of interpretation of the theories concerned. In particular, an example like the relation of atomic physics to chemistry involves the transition from quantum to classical physics; and so understanding it is bedevilled by the measurement problem! In view of these complications, and perhaps controversies, we must expect the 'higher' levels to have much (e.g. in terms of structures, concepts or explanations) that is distinctive, giving them a lot of 'autonomy' from the 'lower' levels.

Of course, this is not the place to investigate such relations. (I myself would be happy to describe the higher level as 'emergent' from lower levels, not as reduced to them; but as it stands, 'emergent' is too vague, connoting just some relation of dependence weaker than reduction. In particular, I am suspicious of making emergence precise in terms of supervenience; cf. Section 2 of Butterfield and Isham (1999).) But theologians and other 'outsiders to physics' should bear in mind that to the extent that higher levels have such autonomy, the less pressure there will be on them to take their lessons in ontology from physical theories of the lower levels.

(2) My second remark concerns approximation in the relation between theories. In general, the idea that a 'higher' level or theory might only approximate an underlying 'lower' level is just one way among others that, as just discussed, inter-theoretic relations can be subtle and complicated. But such approximations will also play a significant role in what follows: we will see various cases where in trying to relate the classical and quantum levels, all one can reasonably demand is that the classical should approximate the quantum. The details of the notion of approximation will vary from case to case. In some cases, such as decoherence (Section 4), it is a matter of numerical predictions by classical theories being close to those of quantum theory; in other cases, such as Everettian interpretations (Sections 6 et seq.), it is a matter of classical concepts being used as approximations by which to grasp novel quantum concepts.

### 3. The Measurement Problem

I turn to expounding the measurement problem of quantum theory (hereafter, QT). It arises from a central feature of QT's success in describing the microrealm of electrons atoms etc.: namely, its denial that these objects always have definite values for quantities like position, momentum and energy. In short, the problem is that QT apparently implies that this indefiniteness should also be endemic in the macrorealm, i.e. the familiar realm of tables and chairs. This is called the 'measurement problem', mainly because the argument that QT implies an indefinite macrorealm is clearest for a measurement situation. Thus, if you use QT to analyse a measurement of, say, the momentum of an electron, which QT says has no definite momentum, you find that according to QT, the indefiniteness of the electron's momentum is transmitted to the apparatus' pointer -- so that it has no definite position.

The details are as follows. Like other physical theories, QT assigns states to systems: the state fully specifies the properties of the system. But the orthodox interpretation of a quantum state is as a catalogue of probabilistic dispositions ('Born-rule probabilities'). That is: for each quantity (position, energy, momentum etc.), the state defines a probability distribution on all possible values of the quantity. These states fall into two kinds.

First, there are 'pure states', represented by vectors and written (in Dirac's notation) with angle-brackets, e.g. $|\psi>$. These vectors are equipped with notions of length and angle. So it makes sense to say that $|\psi>$ is of length 1, or that two vectors are orthogonal ('at right angles'); and more generally, the usual notions about vector spaces, such as 'basis' and 'orthonormal basis', apply. The set of vectors representing the possible pure states of the system is called the 'Hilbert space' of the system. For each pure state, there are some physical quantities and some value of each such quantity, such that: the state ascribes probability 1 to that value for that quantity. But for each state, the great majority of quantities are ascribed a non-trivial probability distribution. To explain this, we first need some mathematical notions and jargon.

Each quantity is represented as a certain kind of function ('operator') on the system's Hilbert space (viz. a self-adjoint operator); and if a quantity sends a vector to a real-number multiple of itself, then we say that the vector (or state) is an 'eigenvector' ('eigenstate') of the quantity, and the real number is the corresponding 'eigenvalue'. Given a quantity, most vectors are not eigenvectors of it, but rather sums of eigenvectors (with different eigenvalues from each other); such sums are called 'superpositions'. The set of eigenstates of a given quantity, for a given eigenvalue, is called an 'eigenspace': it forms a subspace of the space of all vectors, just as a plane through the origin forms a subspace of ordinary three-dimensional space. If the subspace is one-dimensional (like a ray through the origin), the eigenvalue is called 'non-degenerate'; if it is many-dimensional, the eigenvalue is 'degenerate'.

It turns out that every quantity is 'built' in a simple way from a special kind of quantity, called a 'projector'; (and so discussion often focusses on projectors). For any subspace, the function sending all vectors orthogonally into the subspace (so that each vector in the subspace is sent to itself) is the projector onto that subspace. The subspace is the projector's range, and is also its eigenspace for eigenvalue 1; and the subspace of all vectors orthogonal to the given subspace is its eigenspace for eigenvalue 0. So projectors have just two eigenvalues, 1 and 0.

It turns out that every quantity is a linear combination (with real-number coefficients) of projectors onto mutually orthogonal subspaces, so that these subspaces are the quantity's eigenspaces and the real-number coefficients are its eigenvalues. The projectors onto these subspaces are called the quantity's 'eigenprojectors' or 'spectral projectors'. This implies that given a quantity we can find a set of its eigenvectors that are of length 1, mutually orthogonal and together form a basis (implying that any vector is a linear combination of them). This set is an orthonormal basis; since it consists of eigenvectors of the quantity concerned, it is also called an 'eigenbasis' of the quantity: but I shall usually just say 'basis'. Furthermore, if every eigenprojector of a quantity is one-dimensional (i.e. its range is a one-dimensional subspace)---equivalently, if every eigenvalue of the quantity is non-degenerate---then the quantity's eigenbasis is essentially unique, and the quantity is called 'maximal' (or 'non-degenerate'). But if an eigenvalue is degenerate, there is freedom in specifying an eigenbasis: for one can rotate a set of orthogonal eigenvectors spanning the eigenvalue's eigenspace.

Turning to the physical interpretation of vectors as states that prescribe probabilities for quantities: an eigenstate of a quantity ascribes probability 1 to the corresponding eigenvalue. This is a special case of the general rule whereby the state's probability distribution for a quantity is derived from the geometry of the vector space. In the general case, where the state is a superposition of the quantity's eigenstates, the probability prescribed by $|\psi>$ that quantity Q takes value q is calculated as the square of the length of the vector got by projecting $|\psi>$ down in to Q's eigenspace for eigenvalue q. The convention that the original $|\psi>$ is of length 1, secures (by a simple generalization of Pythagoras' theorem) that the probabilities for the various possible values add up to 1.

The second kind of state is more general: these are the 'mixtures', represented by a weighted sum of pure states, the weights adding up to 1. For any quantity, they ascribe as its probability distribution the weighted sum of the corresponding distributions ascribed by the pure states. Accordingly, mixtures are useful for calculating statistics for a heterogeneous population ('ensemble') of systems, in which different pure states occur in different proportions. A mixture is conveniently represented as a special kind of quantity, i.e. a special kind of self-adjoint operator on the Hilbert space of vector states. Namely, it is represented as the linear combination of the projectors onto the rays (1-dimensional subspaces) that contain the mixture's component pure states (with the coefficient of each projector being its corresponding weight). Being a linear function on the vectors, this representation can be expressed as a matrix, relative to any basis of the Hilbert space; this is called the 'density matrix'.

But mixtures cannot always be interpreted straightforwardly, in terms of a heterogeneous ensemble of systems (called 'the ignorance interpretation'). For, surprisingly, for any composite system, the vast majority of its pure states (namely the 'entangled' or 'non-factorizable' pure states) determine as the state of each component system, a mixture. That is: for any entangled composite state, there is a set of pure states and associated weights for the component system such that the correct probability distribution for any quantity on the component system is given by the weighted sum of the distributions ascribed by the pure states. But beware! These mixtures cannot be given the ignorance interpretation, on pain of forcing the composite system to be in a mixture. That is: if we have an ensemble of composite systems in an entangled state $|\psi>$, we cannot think of each single specimen of the component system as in one the pure states in the mixture determined by $|\psi>$. For if it were, then the composite system would also be in a mixture; (just because a population of composite systems must be heterogeneous if a factor-population of component systems is). These curious mixtures are usually called 'improper mixtures'; and ignorance-interpretable mixtures, i.e. mixtures representing a heterogeneous population, are called 'proper'.

For a given quantity, a superposition of its eigenstates and a mixture might ascribe the same non-trivial distribution. But pure states and mixtures are always distinct: any superposition, and any mixture, will ascribe different probability distributions from one another, for some quantity or other -- perhaps a very arcane one. The numerical differences in the distributions, value for value, are 'interference terms'.

So far, I have just summarized the formalism. Matters become more contentious when we ask: what exactly are these probabilities of? The orthodox answer is that they are probabilities of measurement results. There are two aspects to this orthodoxy: the first restricts when a system has a value, while the second concerns measurement.

(1): It is orthodox to say that the system has a value for a given quantity only when its state ascribes probability 1 to that value; (the 'eigenvalue-eigenstate link'.) Agreed, to a philosopher of probability, that will seem a howler. And it may well be wrong. But traditionally, it has been supported by two considerations. Firstly, supposing there are values apart from those ascribed probability 1 ('hidden variables'), it has seemed impossible to design measurements that yield them reliably as results. Secondly, some proposals for such values, though simple and natural from a formal point of view, cannot work: they lead to outright contradiction (as in the theorems of von Neumann, Gleason, Kochen & Specker). And if we evade such contradictions by making the hidden variables 'contextual', still they will need to be non-local in some sense, in order to recover QT's mysterious correlations between entangled but spatially separated systems (as in Bell's theorem).

(2): More generally, orthodoxy says: the probability distribution ascribed by the state is not for values possessed at some time or other; but for values yielded as result (pointer-reading) on a measurement apparatus. It is curious for the interpretation of state to invoke something so extrinsic to the system; and furthermore, for a typical measurement on a typical quantum system, something so vast compared with the system, so variable from occasion to occasion, and so vague. Indeed, it is surely more than curious: it is unsatisfactory, unless you have some general operationalist or instrumentalist view of theories.

This predicament prompts one to analyse measurement processes from the point of view of QT, if necessary in a piecemeal way; hoping to vindicate this interpretation of state. (This endeavour will also form part of a more general project: recovering classical physics, or rather its successful predictions, for the macrorealm, in appropriate limits.) The obvious topic is reconciling quantum theory's scarcity of values with the success of classical physics' opposite assertion, that every quantity has a value on every system.

And thus we arrive at the measurement problem: QT seems to veto such a reconciliation; and to do so in a way that commits it to a manifestly false prediction. For the quantum theoretic analysis of the interaction of, say, an electron that is indefinite for momentum, with an apparatus for measuring momentum, suggests that the electron's indefiniteness will be transmitted to the apparatus -- so that its pointer is in no definite position! This suggestion has been proven for a wide range of exact quantum theoretic models of measurement. But we can confine ourselves to a very simple model.

Before entering the details, I should mention the famous 'projection postulate'. It turns out that many measurements are, to a good approximation, repeatable in the sense that, whatever the initial state of the measured system, an immediate repetition of the measurement always gives whatever result was first obtained. Taken together with both aspects of the above orthodoxy, this motivates the projection postulate ('collapse of the wave packet'): that in a measurement, the measured system goes into the eigenstate corresponding to the result obtained.

As we shall see in a moment, the same line of thought also motivates a similar collapse for the apparatus, to an eigenstate of pointer position. Here I should note a technical point: that position has no exact eigenstates, only approximate ones. But in this discussion, I can ignore this point, which is independent of whether the system considered is microscopic, such as an electron, or macroscopic, like a pointer: it depends just on the fact that position's possible values are a continuous, not discrete set, of real numbers. More important is the fact that it is only in the collapse of the wave packet that the alleged indeterminism of QT occurs: the state of an isolated quantum system evolves deterministically and continuously, according to the Schrödinger equation ('unitary evolution').

So consider a momentum measurement on an electron in a superposition of two momentum eigenstates: one for 1 unit of momentum, and the other for 2 units of momentum. Suppose we have a measurement apparatus or pointer, with 'ready state' $|r>$, which reliably reads these eigenstates, in the sense that the composite system behaves as follows:

$$|1>|r> \to |1>|\text{reads '1'}> \quad \text{and} \quad |2>|r> \to |2>|\text{reads '2'}>.$$

In words: suppose that if the composite electron+pointer is begun in the state on the left of each displayed evolution, then it evolves by the Schrödinger equation in some fixed finite time to the state on the right. Then it follows (by the linearity of the Schrödinger equation) that measuring an electron initially prepared in a superposition yields (ignoring irrelevant normalizations):

$$\{|1> + |2>\}|r> \to |1>|\text{reads '1'}> + |2>|\text{reads '2'}>.$$

But the final state on the right is not an eigenstate of pointer position; (in fact this final state determines the pointer state to be an improper mixture). So orthodoxy, more precisely the eigenvalue-eigenstate link, declares that the pointer has no definite position!

There are clearly two main ways to solve this problem. Either we somehow justify the collapse of the wave-packet for the pointer, so that we replace the above final state by an eigenstate of pointer position; or we somehow supplement the eigenvalue-eigenstate link's meagre ascription of values, i.e. we keep the above final state but nevertheless ascribe a definite position to the pointer (without violating the 'no hidden variables' theorems mentioned above).

In Section 5, I will formulate these alternatives more sharply. In particular, we need to allow, in line with the remarks at the end of Section 2, that a solution to the measurement problem need only recover classical physics' assertion of definite values as an approximation. For example, the first alternative might only secure that the final state is approximately an eigenstate of pointer position; and the second alternative might ascribe a definite value, not to pointer position, but to another quantity that is in some suitable sense 'very close' to pointer position. But to formulate these alternatives more sharply, I first need to express the measurement problem in terms of our two kinds of states, pure states and mixtures; and to discuss decoherence (Section 4).

As presented so far, you might reply on behalf of the orthodoxy that since QT is meant to be indeterministic, and the initial electron state was a superposition, thus ascribing to no result a probability of 1, the final state on the right should *not* ascribe to the pointer a definite position.  Rather, you reply, it should merely ascribe to the pointer, probabilities for the two possible positions, with probabilities given by the electron's initial superposition -- and the final state on the right surely does just that.

This reply on orthodoxy's behalf is coherent, until the last step.  There the reply slides from what it has, to what it needs in order to have a definite result.  That is, it slides: from (a) orthodoxy's official interpretation of the final state, namely that a *subsequent measurement on the pointer* would yield the results, 'reads '1'' and 'reads '2'' (with probabilities given by the electron's initial superposition); to (b) what orthodoxy needs, in order that the pointer *now has* one of the two positions (with probabilities given by the electron's initial superposition) --  namely a proper (i.e. ignorance-interpretable) mixture of two position eigenstates.  But orthodoxy cannot get such a mixture: the final state is pure.

Thus we can summarize the measurement problem as the problem of justifying replacing a pure state by a proper mixture; or in other words, as the problem of eliminating the interference terms that distinguish such states. This summary also leads into our next topic.

## 4. Decoherence

There is a fallacious solution to the measurement problem (often repeated in the textbooks!); which however bears on alot of significant recent work on decoherence. The idea of the solution is to use the fact, noted in Section 3, that if the pure state of a composite system is entangled, then it determines as the state of each component system, a mixture. This seems to solve the measurement problem.  For indeed, the final pure state above determines as the state of the pointer just the right mixture of two position eigenstates!

This solution fails for two reasons.  The first is insurmountable.  It is clear from Section 3: this mixture for the pointer is *not* ignorance-interpretable--it is an improper mixture, not a proper one.  So although it gives the right statistics for subsequent measurements on the pointer, it does not secure that there is a definite result (definite position) at the end of each individual electron measurement.  Again, we see that to get a definite result, we need to go beyond orthodoxy: either by justifying the collapse of the wave-packet, or by ascribing extra values.

The second reason why this solution fails is that its apparent success---determining for the pointer the mixture of two position eigenstates---is an artefact of our simple model of measurement: more specifically, of our measurement's not disturbing the electron's momentum eigenstates.  If it did disturb them, the final pure state of the composite system, that would be obtained after measuring an electron in a superposition for momentum, would *not* determine for the pointer the mathematically right mixture of two position eigenstates.

(I say 'mathematically' since for the first reason above, such a mixture is not interpretatively right---it is improper.) In fact, the final pure state would determine a mixture with the right weights, but of two eigenstates of some quantity different from position -- which quantity depending on how the electron was disturbed.

But this second reason *is* surmountable; and work on decoherence shows how. (This is work by many physicists over the last 25 years. Giulini et al. (1996) is a recent monograph; cf. also Berry (this volume).) The central idea is to take account of the fact that the pointer is not an isolated system: it is continually interacting with its environment of e.g. air molecules or photons. In the physics of this interaction, the pointer's position is crucial: and this enables us to argue, for many plausible models, that no matter how much the measurement disturbs the electron, the final state of the 'total system', electron+pointer+environment, determines for the pointer a mixture very close to the (mathematically, though not interpretatively) right mixture of position eigenstates. The process of obtaining this mixture is called 'decoherence'. (The name arises from the fact that quantum theorists often use 'coherence' to describe phenomena that manifest interference terms, i.e. the state's being a superposition rather than a mixture.) Here, 'very close to' means, roughly speaking, that the eigenstates that are components of the mixture are close to position eigenstates; where eigenstates are said to be close if they are close as vectors in Hilbert space (or in physical terms: if they ascribe very similar probability distributions to any quantity). A bit more precisely: the density matrix that expresses the final state in terms of position is nearly diagonalized; i.e. its off-diagonal terms, often called 'tails', are nearly zero.

So in effect, pointer position gets back 'into the act', not because of its initially intended role (registering the electron's momentum), but because of its importance in the physics of the pointer-environment interaction. And accordingly, the argument applies not only to apparatuses' pointers, but also much more widely: to all kinds of macroscopic systems ('macrosystems') that have a small number of macroscopic or slow or massive degrees of freedom (like the position of the centre of mass), that interact with an environment with very many microscopic or fast or light degrees of freedom such as air molecules or photons. The macrosystem can even be tiny, e.g. a dust particle $10^{-3}$ cm. in radius. In short, plausible models of many such cases lead to the macrosystem state being very close to a mixture of position eigenstates.

Furthermore, the decoherence process is extremely efficient. That is: in many plausible models, the macrosystem's state becomes amazingly close to the right mixture, amazingly fast. Thus Joos & Zeh (1985) consider (among other examples) a dust particle of radius $10^{-3}$ cm. in air. They consider the interference terms distinguishing a superposition, of two positions for the (centre of mass of the) dust particle, just $10^{-4}$ cm. apart, from the corresponding mixture. They show that these terms converge to 0 like

exp – ( t/$10^{-36}$ sec.) ! And they stay very tiny for very long time-scales ($10^{10}$ years).

So what do these results imply about how to solve the measurement problem? I think there are three points, (A)-(C) to be made at this stage: the ubiquity of decoherence means that they concern macrosystems in general, not just apparatuses' pointers.

(A) The first point is a warning. The above description of decoherence is simplified. In particular, a more exact description would allow for the following four points. (However, I will not need to take notice of these points until I discuss the Everettian strategy in Section 6 et seq..)

(i) There is in general no natural unique definition of 'the macrosystem' to be considered, since there will be rapid and efficient decoherence for any of a wide range of definitions. For example, the dust particle rapidly decoheres from its environment, for a wide range of choices of which molecules to include in it.

(ii) Similarly, there is in general no natural unique definition of the relevant 'final state', since the decoherence type of interaction does not get 'turned off': it continues---and continues to keep the interference terms (the 'tails') small.

(iii) As mentioned in Section 3, there are only approximate eigenstates of position. Besides, one can argue that to recover a definite macrorealm one needs to obtain not just any such approximate eigenstates of position, but some that are well-behaved as regards other quantities such as momentum, in particular by evolving according to approximately classical laws of motion. Indeed, there are such states ('coherent states'); and they can be obtained by decoherence, in the sense that in some models, the components of the macrosystem's mixture are very close to coherent states.

(iv) Decoherence does not provide (a) a *basis* of states of the macrosystem, nor even (b) a precise specification of a non-maximal quantity on it, i.e. a mutually orthogonal family of many-dimensional subspaces. For (a): specifying a non-maximal quantity, say the position of the centre-of-mass of a pointer, does not specify a basis, for it leaves open a great deal about the other degrees of freedom of the pointer (including e.g. the positions of its individual atoms). Point (b) is a generalization of (ii) above: the requirement that interference terms be small (or be soon made small, and stay small etc.) is of course vague, and so singles out a vague range of quantities on the macrosystem.

(B) Second: for all decoherence's efficiency, the mixture for the macrosystem that it provides is still not 'interpretatively right': it is improper. This feature is not at all affected by taking cognizance of (A)'s 'wrinkles' (i)—(iv). Nor is it contentious: some authors suggest that we should speak, not of 'decoherence', but of the 'diffusion (into the environment) of coherence'. So: to solve the measurement problem, we must still choose between the alternatives at the end of Section 3: either justify the collapse of the wave-packet (applied to macrosystems), or ascribe extra values.

(C) But third: decoherence results are clearly relevant to the solution. Exactly how they are relevant will vary from one solution to another, but we can already make two general points---which will be particularly relevant to my discussion of the Everettian strategy.

(i): The results exemplify the remarks at the end of Section 2, that a solution to the measurement problem need only recover classical physics' assertion of definite values as an approximation. For the astonishing numbers in these results liberate us from the traditional aim of getting macrosystems to have definite values for *exactly* the familiar quantities like position. Indeed, there are two possible lessons here: one uncontentious, and one contentious. The distinction between them will become clearer in Section 5's discussion of (DefMac) vs (DefApp); and it will be important in Section 7's discussion of Saunders and Wallace.

(a): Once we see how rapid and ubiquitous the decoherence process is, we realize we should be willing to accept a solution to the measurement problem that claims that rather than position itself, some other extremely close quantity is definite in value; (where 'closeness' of quantities can be made precise as indicated above, in terms of closeness of the quantities' eigenstates). For surely nobody can be so certain that *exactly* position, as against a 'close cousin', is definite. (b): Maybe we should go further, and be willing to accept a solution that does not even specify some single quantity as definite-valued, but instead just states a vague range of quantities, the definiteness of any one of which would give a definite macrorealm.

(ii): The results show us that if we endorse Section 3's second alternative, and more specifically propose to ascribe extra values to position for macrosystems, then orthodox quantum theory on its own brings us close to our goal. Roughly speaking, we have only to go a bit further, from an improper mixture to a proper one; and this will in effect be what the Everettian strategy proposes (Section 6 et seq.).

## 5. Two Choices Yield Four Strategies

I turn to surveying strategies for solving the measurement problem. We have seen that one has to choose between justifying the collapse of the wave-packet, and ascribing extra values to quantities. Let us make this a little more precise, and introduce labels. We have to choose between:

(Dynamics): We secure a definite macrorealm by denying the Schrödinger evolution of a strictly isolated system, instead proposing new dynamical laws that can evolve a pure state (superposition) into a (proper, i.e. ignorance-interpretable) mixture.

(ExtraValues) We keep the Schrödinger evolution of a strictly isolated system, and solve the measurement problem by ascribing to certain quantities some values beyond those prescribed by the eigenvalue-eigenstate link (while avoiding the various 'no hidden variable' proofs).

But there is also another choice to be made, which is independent of this first one; so there will be four broad strategies to choose from. As I have described the measurement problem so far, I assumed that the macrorealm was definite: i.e. that macroscopic objects have definite values---if not for position, then at least for very 'close' quantities. But on reflection, this assumption is debatable.

For one can take the view that the evidence in its favour (the empirical success of classical physics with its assumption of definite values for all quantities) ultimately boils down to our *experience* of the macrorealm being definite. (One might retort that this view depends on some sort of foundationalist epistemology: which are now largely discredited. But let us follow the line of argument.) So perhaps one should retrench, and in stating the measurement problem, assume only the disjunction, that the macrorealm is definite---or at least we experience it as definite. But in that case, the requirement on a solution to the measurement problem becomes disjunctive: either to secure a definite macrorealm, or to explain why it seems definite to us.

It is worth putting this point in terms of the general scientific enterprise of accounting for our experience of the world. Hitherto, our discussion of the measurement problem has envisaged that once a definite macrorealm is secured, an account will then be given of why we experience it as definite. The discussion can admit the undeniable fact that such an account will be very complicated; (as is shown by the complexity of theories in perceptual psychology). And no doubt, the account must eventually address controversial issues about the scientific understanding of consciousness.  These complexities and controversies are considerable. But the present point is that so far, there seems to be nothing quantum-theoretic about them. That is to say, one naturally envisages that once a definite macrorealm is secured, some sort of non-quantum account (no doubt rooted in classical physics and physiology) will 'take over the job' of accounting for experience. On the other hand, once we entertain the idea that the macrorealm is indefinite, but that somehow or other, we nevertheless experience it as definite---one naturally envisages that there will be some kind of quantum-theoretic weirdness about the account of experience.

Again, it is worth introducing labels for the two alternatives:

(DefMac): We solve the measurement problem by securing a definite macrorealm; and expect some sort of 'classical psychophysics' to account for experience, and in particular to explain why the macrorealm also *appears* to us to be definite;

(DefApp): We allow an indefinite macrorealm; we solve the measurement problem by securing only that it *appears* definite---and so expect some sort of 'quantum psychophysics'.

This second choice is independent of our first one. All four combinations are logically coherent strategies for solving the measurement problem; and indeed, all are advocated in the literature. Of course, since the two choices and ensuing four strategies are cast in broad terms, there will be a considerable variety of views under each strategy; and since the strategies are cast in rather vague terms, some specific views could perhaps be as well classified under another strategy.

The rest of this paper will concentrate on the Everettian version of the fourth strategy, that combines (ExtraValues) and (DefApp); and in particular on an example of this strategy articulated by Saunders and Wallace. (The point just made applies here: the underlined labels being vague means that some Everettians, Saunders and Wallace included, could perhaps be as well classified under (DefMac).) But in this Section, I shall briefly mention, and give some references for, the other three strategies. Surveying these strategies and views will help locate the Everettian strategy in a landscape of 'possible ontologies' for quantum theory.

First, (Dynamics) and (DefMac). The obvious examples of this strategy are the proposals of such authors as Gisin, Ghirardi, Pearle and Percival; all of whom propose precise equations for the 'collapse of the wave-packet' of an isolated system, in such a way as to secure at the end of a measurement procedure, an ignorance-interpretable mixture with the correct (Born-rule) probabilities for its components. (Cf. Ghirardi et al. (1995) for an exposition emphasising interpretative issues.)

These proposals raise many interesting issues, both physical and interpretative; (though these two kinds of course overlap!). Prominent among the physical issues is the question: What are the prospects for relativistic models of this kind? So far the best-established models are non-relativistic. Prominent among the interpretative issues is the treatment of 'tails'. That is to say: in these models, the (pure) quantum state of an individual system at the end of a measurement is not an exact eigenstate of the preferred quantity that according to the model 'controls the collapse'. The state has 'tails', i.e. small non-zero components for other values of the quantity. Is this a defect? Or is it enough to say what I said about decoherence at the end of Section 4: that we simply learn that it is not exactly the quantity we originally preferred in setting up our model, that has definite values? (I myself answer No and Yes to these questions.)

I turn to (Dynamics) and (DefApp). The obvious examples of this strategy are the proposals of Wigner (1962) and more recently Stapp (1993), that mind (or consciousness) itself produces the collapse of the wave-packet. These authors do not have well-developed models of the collapse; (though Stapp has 'toy-models'). But the idea is that the usual Schrödinger evolution holds throughout the physical realm, and is broken only at the interface of brain and mind: once the mind sees one measurement result, the measured system's (and the pointer's) superposition is replaced by an eigenstate, namely the one corresponding to the result seen.

I admit to giving this sort of proposal very little credence (Butterfield 1995, pp. 130-131; 1996a, pp. 151-156): (though it has the merit of instantiating my general theme, that we should be willing to be radical in interpreting quantum theory!). But I should explain why in my taxonomy, these proposals exemplify (DefApp), not (DefMac); since doing so will clarify a contrast with the Everettian strategy.

I admit that according to this proposal, once consciousness has 'done its stuff' and collapsed the wave packet, the macrorealm really is definite (for the quantities of which the collapsed state is an eigenstate). That is: the components of the initial superposition corresponding to the values that are not seen by consciousness, *really* disappear. It is not just a matter of it being utterly impractical, thanks to decoherence, for us to confirm their existence by experiments whose statistics manifest the interference terms. Nevertheless, I classify these proposals as examples of (DefApp). For this definiteness of the macrorealm is in general limited and temporary. After 'consciousness stops looking', the usual Schrödinger evolution takes over again. So there is no guarantee that the macrorealm stays definite (in the favoured quantities) as times goes on, in particular when consciousness stops looking—'when there's no one about in the Quad'. This temporary definiteness of the macrorealm marks a major difference from the Everettian strategy; as we shall see in subsequent Sections, it takes the definiteness of the macrorealm to be *always* an illusion.

So much for strategies that adopt (Dynamics). From now on I consider those that adopt (ExtraValues). Any such strategy must address two related questions about its postulated extra values: how they solve the measurement problem (either by (DefMac) or by (DefApp)), and how they evolve in time. Though I will emphasise the first question, the second is also crucial: in particular, the values must not evolve so as to undermine later on the solution to the measurement problem that they provide at some initial time!

I turn to (ExtraValues) and (DefMac). The obvious examples of this strategy are the pilot-wave interpretations initiated by de Broglie and Bohm between the 1920s and 1950s, and much developed since then; (cf. Bohm & Hiley (1993), Holland (1993); and especially for historical and philosophical issues, Cushing (1994)). The main ideas of these interpretations are: a quantity is " preferred *a priori*" in that it always has a definite value; the preferred quantity is position in elementary quantum theory, and field configuration in quantum field theory; the Schrödinger evolution of the quantum state prescribes a deterministic evolution of this quantity's definite value; and this definite value solves the measurement problem (at all times!). Since Cushing (this volume) will give more details of these interpretations, I will only expand a little on my remark in Section 1, that they are perfectly tenable: again, doing so will clarify a contrast with the Everettian strategy.

The first point to make is that there are pilot-wave interpretations which can plausibly claim to be empirically adequate: i.e. to recover the well-confirmed empirical predictions of quantum theory---and even relativistic quantum theories, including quantum field theories.  However, they lack a feature that physicists tend to value highly, namely fundamental (as against phenomenological, or average) Lorentz-invariance: a feature which the Everettian strategy promises to retain.

But this is not say that the Everettian strategy has an indisputable advantage. The pilot-wave theorist can reply that quite apart from empirical adequacy, their interpretations also have some valuable *theoretical* features, in addition to being deterministic (a traditional desideratum for fundamental physics). For example, they give a wholly lucid and  unproblematic modelling of quantum theory's curious non-local phenomena as arising from an action-at-a-distance that is no more incomprehensible than Newton's gravity; though of course there are differences from Newtonian theory, such as that the force's strength does not decrease with distance.  But I do not need to try to adjudicate between the Everettian and pilot-wave theorist. It is enough here to note, in line with Cushing (this volume), that while rejecting pilot-wave interpretations (say because one thinks the lack of fundamental Lorentz-invariance is a disadvantage weightier than all their advantages) is legitimate---one should be clear that one is doing this on grounds that go beyond the empirical.

I should add that other examples of the (ExtraValues) and (DefMac) strategy include the modal interpretations proposed in the 1980s by such authors as Bub, Dieks, Healey, and Kochen. In fact, there are connections between these and the pilot-wave interpretations; cf. Bub (1997). Whether these interpretations can succeed as regards such issues as empirical adequacy and Lorentz-invariance is still an open question. The 'state of the art' is represented by some recent books: Dieks and Vermaas (1998), and Vermaas (2000).

### 6. The Everettian Strategy Introduced

Finally, I turn to the fourth strategy, on which the rest of the paper will focus: (ExtraValues)  and (DefApp). So the idea will be to keep the Schrödinger evolution of a strictly isolated system, and to allow an indefinite macrorealm; but to posit extra values so as to secure the appearance that it is definite. As we shall see, we can think of these values as in some sense perspectival (or even 'mental').

This strategy, pioneered by Everett (1957), encompasses several different interpretations; (Barrett (1999) is a fine survey). I have elsewhere discussed some of them (under the titles 'Many Worlds' and 'Many Minds'; 1995, 1996). I will not repeat any of that discussion here, but focus instead on another interpretation elaborated by Saunders, which he calls 'relativism' (or 'relational QT'; 1995, 1996, 1996a, 1998), and a related interpretation of Wallace (2001, 2001a). As I mentioned in Section 1, Deutsch, Gell-Mann & Hartle, Vaidman and Zurek have advocated similar interpretations; cf. in particular Vaidman (1998, 2000).

I have not the space to discuss in detail these other five authors. But for clarity, I should say at the outset that all seven authors share the following two main proposals---with which some other Everettians would disagree. (1): They secure the appearance of a definite macrorealm by appealing to decoherence to specify appropriate states of macrosystems (including tiny dust particles). (2): They accept with equanimity that vagueness about decoherence, e.g. in saying that an interference term is small (how small? how soon?), makes for vagueness about what is definite. The idea here is as at the end of Section 2, and in (i)(b) at the end of Section 4. Namely: in securing definite appearances, it is perfectly reasonable to resort to approximate, indeed anthropocentric or perspectival, notions. In particular, it is enough to specify a vague range of quantities, the definiteness of any one of which would give definite appearances.

It will be clearest to first explain some basic ideas about the Everettian strategy, ideas that are common to all the interpretations it encompasses. That will occupy this Section; the next two Sections will distinguish between such interpretations as 'Many Worlds', 'Many Minds' and relativism.

At its simplest, the Everettian strategy is as follows. One assumes that the entire universe has a quantum state, presumably a pure state $\Psi$ say; (but the discussion to follow could be generalized to the universal state being a mixture). $\Psi$ always evolves according to the Schrödinger equation: after all, the universe is the only truly isolated system! $\Psi$ can of course be expressed as a superposition of states in myriadly many ways. But the measurement problem suggests it will be a superposition corresponding to many different definite macrorealms. As stated, this suggestion is of course vague, since 'definite macrorealm' is vague. But the measurement problem suggests that however one makes it precise or at least more precise (e.g. by appealing to decoherence), $\Psi$ will have a component (i.e. a summand with non-zero coefficient) corresponding to many different definite macrorealms (and presumably many other states too). At this point, the Everettian proposes their breath-taking main idea: that all these definite macrorealms are actual.

This main idea needs a lot of clarification; indeed Everettian interpretations differ so greatly in how they clarify it, that the common slogan 'the various definite macrorealms are actual' is very ambiguous. I will first give four clarifications that apply to all versions of the Everettian strategy. Then in Sections 7 and 8, I will distinguish different Everettian interpretations. Section 7 concerns how Everettians use Everett's idea of relative states to define their 'extra values', i.e. 'definite macrorealm'; and Section 8 concerns the treatment of time, including the evolution of macrorealms over time. Though I think this order is optimal, it is not perfect, since the various aspects of the discussion are of course related. In particular, postponing the treatment of time to Section 8 means that until then, we lose sight of: (i) the desideratum that 'macrorealm' should be so defined that the macrosystems they contain are suitably stable over time; (ii) one of the principal claimed advantages of the Everettian strategy---that (unlike for example pilot-wave interpretations) it allows a fundamentally relativistic dynamics.

(Actual): First, I should relate these different but actual macrorealms to the standard philosophical framework of possible worlds, in Leibniz and modern modal metaphysicians. Recall that each such world is a logically possible total course of history; one such world is actual; propositions, including laws of nature (if such there be), are made true at worlds. Let us assume from now on (with touching faith!) that the actual world obeys the laws of QT, including the usual Schrödinger equation. More exactly: the actual world throughout all time (the total course of world-history) is described by the sequence of states, $\Psi(t)$ for all t. This assumption simply sets aside QT-violating worlds, which are irrelevant for investigating the 'world-picture' of QT. Then the basic idea above is that the actual world contains very many macrorealms: all definite, e.g. for the position of some pointer; but different from one another, e.g. as regards which position the pointer is in. So according to philosophers' standard usage of 'world', the name 'Many Worlds' is a misnomer.

I should add that some Everettians including Saunders prefer a different usage. They say that the different macrorealms are all 'real', and they reserve 'actual' for the macrorealm " we are in" (in a sense yet to be clarified!). I agree that this conforms at least equally well with everyday usage, but I have preferred to align my use of 'actual' with philosophers' current use of 'actual world'. (Some Everettians also use 'multiverse' for the one actual quantum cosmos, containing many macrorealms or 'worlds'.)

(Branch): The second comment follows on from the first. You might object that for a pointer to have different values of its position -- more generally for a system to have two or more values for a quantity -- is downright impossible. So such a constellation of different macrorealms cannot be in a single possible world. But to this, the Everettian replies that she is postulating a hitherto-unnoticed parameter, relative to which objects have properties. Contradiction is avoided, since the object has the contrary properties relative to two different settings of the parameter. This line of reply is certainly coherent: it is similar to the familiar way in which philosophers say that change of properties over time involves no contradiction, since the object has the contrary properties relative to two different times. (This analogy with time will be developed at the end of Section 7, and in Section 8.)

Again I should clarify terminology. The literature uses various words, such as 'perspective' and 'branch' as well as 'world', for this new parameter---or rather, for its values; but none is dominant. And none is ideal: as we will see below, 'perspective' can sound too subjective; and 'branch' suggests the metaphor of a tree-trunk splitting into branches, and thereby the idea of indeterministic evolution over time (with the branches representing alternative possible histories)---which might be misleading, as we will see in Section 8's discussion of time. Until then, I shall for the most part use '(definite) macrorealm', 'world' and 'branch' synonymously: each to be understood, until Section 8, as defined only at an instant.

However, I should also note a helpful terminological proposal by one of 'my' authors, Wallace. He proposes to use 'branch' for what I have so far labelled a '(definite) macrorealm'---which notion he (together with Saunders) takes to be: approximate; defined by decoherence; but nevertheless, in part anthropocentric; temporally extended, i.e. not defined only at an instant; and to evolve stably (in some sense of 'stably'). And since he also advocates that an arbitrary resolution of the universal quantum state $\Psi$ is perfectly legitimate (though usually useless), he proposes to use 'world' for the corresponding 'aspects of the multiverse' (i.e. aspects of the philosophers' 'actual world'): so that worlds are precise, defined arbitrarily, and in general defined only at an instant. But beware: though Saunders' and Wallace's views are very close, Saunders tends to resist talk of 'branches' and 'worlds'. Besides, there is even more variety of usage in the literature: e.g. Vaidman uses 'world' for what is roughly Wallace's 'branch'.

(Link): Third: note that *relative to a branch*, the Everettian strategy retains the orthodox eigenvalue-eigenstate link of Section 3. That is: it is precisely because a summand of $\Psi$ is an eigenstate of a quantity e.g. position of some pointer, that Everettians take the branch described by that summand to be definite for that quantity.

(Indefinite): Fourth: beware of an ambiguity. I have said (in Section 5 and at the start of this Section) that the Everettian strategy adopts (DefApp) rather than (DefMac), and so has an indefinite macrorealm. This may now seem odd, since I have just said it involves a plethora of definite macrorealms. There are two points here. The first is that this is a spurious conflict: a mere matter of words. We can either say there is one macrorealm that is indefinite but has perspectives or branches, relative to each of which there is definiteness; or we can say just that there are many macrorealms, each definite. The original formulation of the measurement problem as a threat of indefiniteness prompts the first usage: while the Everettian's main idea prompts the second. In short: to have many different definites (relative to different settings of a parameter) is a way to be indefinite.

The second point leads us into differences among the various Everettian interpretations. It is that among their rival proposals for how to make 'definite macrorealm' (or 'world' or 'branch') precise, some are mentalistic in a more radical sense than Saunders' and Wallace's acceptance that branches be defined in an anthropocentric way. These proposals explicitly invoke a notion of mind (or some very similar notion such as sentience). Hence they are called Many Minds interpretations. This means that they take 'definite macrorealm' (or 'world' or 'branch') to require only that what in fact appears to minds (or sentient beings) be definite. So for example, a macrosystem such as a pointer can be indefinite in position, even *within* a branch. All that is required is that *if* it appears to minds, then it appears to them to be definite in position; and this could be secured by appropriate quantities defined on brains (viz. those quantities that underpin states of perceiving a pointer to have a definite position) taking definite values relative to a branch. So these interpretations endorse (DefApp) in a more radical sense than their rivals. (This will become clearer in the next Section which discusses Everettians' rival proposals for how to make 'definite macrorealm' precise. For the moment, I just note that this conception of the macrorealm as indefinite *within* a branch is obviously analogous to the Wigner-Stapp view discussed in Section 5.)

## 7. Defining the Branches

I turn to address what is perhaps the main question facing the Everettian strategy—how should it define 'definite macrorealm' (or 'world' or 'branch')? This question is of course the analogue for the Everettian strategy of the questions about which quantity is preferred that confront Section 5's other views. Accordingly, it might be called the 'preferred quantity problem'. Unfortunately, it is more often called the 'preferred basis problem', suggesting (wrongly) that the Everettian needs to specify not just a quantity, but a basis; (or equivalently apart from the choice of eigenvalues: a maximal quantity, i.e. a quantity all of whose eigenvalues are non-degenerate). Indeed 'preferred quantity' also suggests that the eigenvalues, not just the orthogonal family of eigenspaces, matter to the Everettian strategy: which is presumably false. So a phrase like 'preferred decomposition' might be even better; but I shall stick with the more familiar 'preferred quantity'.

This is a multi-faceted question, to which Everettians have given very different answers; and any full answer must of course include a discussion of how macrorealms (branches) evolve over time. But it will be clearest to postpone that (to Section 8). For the various Everettian answers make use of a common formal notion, viz. the notion of relative state, introduced by Everett himself; and this notion, and even the main Everettian proposals for how to fill it out, can be explained independently of the treatment of time. (Indeed, the notion is itself needed for the discussion of time.) But even before explaining the notion of relative state, and how different Everettians propose to fill it out, I need to draw attention to a general philosophical issue, about vagueness.

### 7.1 Vagueness Acceptable?

One naturally requires the Everettian to propose a precise definition of 'macrorealm', 'world' or 'branch'; i.e. to specify a preferred quantity (as the pilot-wave and modal interpretations do). For the Everettian's idea is that each of her proposed worlds is actual; and it surely cannot be a vague matter what actuality itself consists of. (In particular, the philosophical orthodoxy about vagueness is that words or concepts being vague is a matter of semantic indecision: we simply have not decided fully what our words or concepts mean---in which case there is little scope for actuality itself being vague.) Yet it turns out that Everettians' definitions of 'macrorealm', 'world' or 'branch', are usually vague (cf. Section 7.2).

So the question arises: is a vague definition somehow acceptable? As mentioned above, Saunders' and Wallace's answer is 'Yes'. To be more precise:-- (i) Saunders is reluctant to talk of worlds or branches, but—at least as I read him---that reluctance reflects his believing them definable only vaguely (partly because they are anthropocentric notions). (ii) In Wallace's usage, the Yes answer pertains to the 'branches', which are defined vaguely because anthropocentrically, but not to the worlds---which are precisely defined. So overall, Saunders reads as more anthropocentric than Wallace. (This is in part a result of Saunders' views being similar to those of Gell-Mann and Hartle: all three authors relate 'branch' etc. to the existence of stable records, and thereby to organisms or other complex macrosystems exploiting such records so as to survive in their local environment---topics I will touch on again later.)

To clarify this answer, I need to formulate two ways in which vagueness is acceptable (the first way being very straightforward). First, vagueness about precise matters is of course acceptable if it is simply temporary: a matter of having a working definition, which one plans to make precise. But one might complain that the Everettians have had some 40 years to make it precise! And one might add that Everettians' intermittent attempts at precise definitions have been found to have flaws: a salient recent example concerns the consistent histories approach advocated by Saunders; cf. Dowker & Kent (1996).

The second way that vagueness can be acceptable arises from a general point about explanation. Namely, if an explanandum is vague, its explanans can be vague. I say 'can be' not 'must be'. I agree that the explanans *could* be precise: if so, it would thereby explain directly one precise version of the explanandum, the other precise versions being explained to the extent that they approximate it.

That is to say: if an explanandum is vague, the explanans can be vague (provided that its vagueness does not somehow undermine its explanatory role). Indeed, such vagueness would be an advantage if it somehow corresponded to the vagueness of the explanandum, in such a way that precise versions of the explanandum were satisfactorily explained by corresponding precise versions of the explanans. This idea of corresponding vaguenesses is familiar in the philosophy of language. Thus Lewis (1973 Chapter 4.2) defends his use of the vague notion of similarity in his proposed truth-conditions for counterfactual conditionals, by maintaining that the kinds of vagueness correspond: resolving the vagueness of these conditionals' truth-conditions amounts to resolving the vagueness of similarity.

These points apply to the Everettian's project of explaining the appearance of the familiar definite macrorealm ((DefApp)). For first: the explanandum at issue is indeed vague---there is no precise agreement about exactly which systems have definite values for exactly which quantities, in a 'macrorealm', nor about what an appearance of such requires. Rather there is a cluster of approximate notions and doctrines relating them. This cluster includes the example I have concentrated on, viz. macrosystems having approximately definite positions. But one can also demand more, such as these positions changing in time in approximate agreement with the laws of classical mechanics. And as I mentioned at the end of Section 6 in connection with 'Many Minds' interpretations, one can instead emphasise the *appearance* of definiteness, and therefore consider a more anthropocentric explanandum. Nor do we *need* to have any precise agreement about this explanandum: recall from the end of Section 2 that often a superseded theory or its empirical predictions are only approximately recovered by its successor.

Second: Everettians' explanans, i.e. their definitions of 'macrorealm', are often vague. As we shall see in more detail in Section 7.2, there have been three main Everettian approaches: one requiring only that (roughly speaking) what in fact appears to minds is to be definite (the 'Many Minds' approach); and two requiring that (roughly speaking) all macrosystems be definite---they differ in that the more traditional of the two approaches takes the vague notions of apparatus and pointer-position as primitive notions, while the more recent approach uses a notion of macrosystem and appeals to decoherence. I shall concentrate on this third approach---to which Saunders and Wallace belong. About the first two approaches, suffice it to say here that they clearly use vague notions (such as mind and apparatus, respectively).

As to the third approach, with its appeal to notions of macrosystem and decoherence, we saw at the end of Section 4 that these are vague notions. (The detailed physics of decoherence is also an active research area, so the first point above, that vagueness can be a matter of working definition, also applies.) In particular, there are the questions about the size of the 'tails' ('how close to diagonal is the density matrix?'), and about which quantity decoherence selects ('in which basis must the density matrix be nearly diagonal?'). And broadly speaking, these kinds of vagueness *do* correspond to different ways of making precise the explanandum. Besides, the ubiquity and astonishing efficiency of decoherence means that for all macrosystems the tails will very quickly be very tiny, and the selected quantity will be very nearly unique---so that the vagueness will be unnoticeable by the standards of precision usual for macroscopic physics.

I conclude that at least on this third approach, the Everettians' vagueness is acceptable; (at least for the moment, while the physics of decoherence is still being explored). In particular, their vagueness corresponds, broadly speaking, to a vagueness in the explanandum; and the proviso I mentioned above---that the explanans' vagueness does not somehow undermine its explanatory role---seems to be met.

### 7.2 Relative States and Preferred Quantities

I turn to relative states. The idea is that for any pure state of a composite system consisting of two component systems, and for any quantity on the first component, and any eigenspace of it, there is a unique associated state of the second component, called 'the relative state' for the given eigenspace on the first component. (The given eigenspace is often taken to be one-dimensional; or equivalently, one considers an element of a basis for the given quantity on the first component. But this restriction is not necessary.) The idea of the 'associated state' can be spelt out most easily, by using the orthodox interpretation of state as prescribing probability distributions for measurement-results; though the idea does not rest on this interpretation. In these terms, 'associated' means that immediately after measuring the given quantity on the first component, and obtaining the result corresponding to the given eigenspace, a measurement of any quantity at all on the second component will have the probability distribution prescribed by the relative state. In short: conditional on the first component's result, the second component 'looks as if' it is in the relative state.

Four points of clarification, in ascending order of importance. (1) The first/second contrast here is of course immaterial: we can similarly define the relative state of the first component, for a given eigenspace of the second.

(2) For readers familiar with QT, it may be helpful to put the idea of relative states in terms of the formalism. I shall for simplicity take the given eigenspace as one-dimensional, i.e. choose a given state on the first component. Recall that any pure state of the composite can be written as a superposition of product states in arbitrary bases on the components, i.e. $\Sigma_{ij} c_{ij} |\phi_i\rangle|\psi_j\rangle$. Given this initial state, we define for the given state $|\phi_i\rangle$ of system 1, the relative state of system 2 to be $\Sigma_j c_{ij} |\psi_j\rangle$ (upto normalization). A calculation shows that this relative state prescribes the same probability distribution for any quantity on system 2, as does the composite system state obtained by applying the projection postulate (Lüder's rule) to the initial state and supposing you get the eigenvalue corresponding only to $|\phi_i\rangle$; (and that this relative state depends on the choice of the pure state of the composite system, and on $|\phi_i\rangle$, but not on the basis of all the $|\phi_i\rangle$s).

(3) I have defined the relative state in terms of the state of the composite system *at a time*, and to that extent the idea of a relative state is not a relativistically invariant notion. But of course relativistic theories are often discussed in terms of the evolution of states from time to time; relativity requires only that there should be a suitable meshing of the state-evolutions associated with two different foliations of spacetime. Besides, as I will mention in Section 8, one can give a foliation-independent generalization of relative states, using the local algebras approach to quantum theory in which quantities are associated directly with spacetime regions: an approach advocated by Saunders.

(4) There is another respect in which the above explanation of relative state might seem restrictive. Because it assumes from the start a composite system with two components, you might think that this decomposition has to be in some way given *ab initio*;

in particular to be independent of the dynamics. Not so: no matter how one defines the composite system and its decomposition into components, the notion of relative state applies. This is important, for a reason often ignored in discussions about interpreting elementary quantum theory. Namely, the particles which these discussions take to compose their systems are, according to our best theory of matter (quantum field theory), excitations of underlying fields; so the very existence of what these discussions take as their systems is, according to quantum field theory, dependent on what the state is. So there are two morals here; first, that the notion of relative states nevertheless applies; and second, that irrespective of this notion, we must in any case accept state-dependent (and so dynamics-dependent) definitions of physical systems.

Let us now see how Everettian interpretations use the notion of relative state, as applied to the universal state $\Psi$, to define their notion of 'world' (or 'branch' or 'macrorealm'). They will differ over: (a) how they split up ('factorize') the Hilbert space for the universe into Hilbert spaces of component systems; and (b) which quantity to select on a given component, thereby defining relative states of the other component.

As I mentioned in Section 7.1, there have been three main approaches. One is naturally called 'Many Minds', since it specifies the preferred components and quantities in broadly mental terms (like 'mind' and 'seeing the pointer to be in a definite position'). I set it aside here, having discussed it elsewhere (1995, Sections 8-10; 1996). The other two are both naturally called 'Many Worlds', since in addressing (a) and (b) they use broadly physical terms (like 'apparatus' and 'pointer-position'). The first, and more traditional, Many Worlds approach assumes a notion of apparatus, i.e. a distinguished set of subsystems of the universe, and a distinguished quantity on such an apparatus, e.g. position of the apparatus' pointer. These assumptions are not linked to decoherence; indeed this approach was developed in the 1960s and 1970s, before work on decoherence took off (in the 1980s). On the other hand, the second, more recent, Many Worlds approach appeals to decoherence to avoid assuming these notions. As we have seen, this means it assumes notions of macrosystem (maybe dynamically defined), and of a dynamically-induced preferred quantity, which is typically a quantity 'close to' position; and the Everettian interpretation that is my main concern, the relativism of Saunders and Wallace, is a version of this last approach.

Agreed, this threefold classification needs to be qualified. For the vagueness of both the mental and the physical terms that are used means that classifying various authors' interpretations is a delicate matter (just as it was for Section 5's three strategies). Indeed, there are two specific points here; of which the second is important for what follows. (1): Although Everett's paper is regarded as initiating Many Worlds interpretations ('Many Minds' being a much more recent phrase), he may well have intended to express a Many Minds view---as several commentators maintain. (2): As mentioned above, some versions of the third approach---including Saunders' and Wallace's version---appeal to anthropocentric considerations in their accounts of macrosystems, and the preferred quantity on them; with the result that in some respects, these versions of the third approach resemble the first, i.e. Many Minds, approach.

In any case, to explain this third approach, I need to first sketch the other, more traditional, Many Worlds approach. It will face two objections, one about vagueness, and one about not securing a definite macrorealm: objections which the third approach can claim to avoid, in part by being willing to appeal to anthropocentric notions.

To sketch the first approach, it will be clearest to focus first on the idealized case of a single apparatus, together with the rest of the universe; then I will turn to the realistic case of many apparatuses. Assuming that there is just a single apparatus, the Everettian proposes that a branch is specified by an eigenspace of pointer-position on the apparatus, together with the relative state (defined by $\Psi$, the state of the universe) of the rest of the universe, for this eigenspace. So, within the branch, the rest of the universe *is* in the relative state; and, using the eigenvalue-eigenstate link within a branch, the rest of the universe has values for exactly those quantities of which the relative state is an eigenstate. (As usual, I am setting aside the fact that there are no exact eigenstates of position.)

As it stands, this proposed definition of a branch seems inadequate. For it seems very unlikely to secure definiteness (in a branch) across the whole macrorealm, for the quantities that appear to us definite. Why should the relative state for my pointer being here, be an eigenstate of my table being just there---and the Moon being way over there?

But the Everettian of course takes the case of a single apparatus to be an unrealistic simplification: in whatever way she makes precise the notion of an 'apparatus', and its special basis 'pointer-position', there will surely be many apparatuses. And supposing there are N apparatuses, she will define a branch as follows. (a) The state-space of the universe is to be factorized into N+1 factors, one factor for each apparatus, and one factor for the rest of the universe. (b) A branch is specified by an N-tuple of eigenspaces of pointer-position on the N apparatuses, together with the relative state of the rest of the universe for this N-tuple: one just treats the N apparatuses collectively as the first component system, in the definition of relative state. (More precisely, not every N-tuple need define a branch: but those N-tuples that have non-zero coefficient in the expansion of $\Psi$ in terms of these N+1-fold products, do so.)

This approach faces various objections. But some of these apply to any Everettian strategy, irrespective of the choice of preferred system and quantity, e.g. objections about the interpretation of probability; and I will discuss some such objections later. And some of the objections that are specific to the approach might well be answered (or are artefacts of my exposition!). For example, (a) and (b) above seem to assume a division of the world in to N apparatuses and 'the rest', independent of the state and of dynamics, contradicting point (4) above. But this approach might well be able to reply, i.e. to accommodate (4), by giving a more sophisticated account of 'apparatus'.

So here, I will just state the two objections which the third approach (including Saunders and Wallace) can claim to avoid. The first concerns vagueness. It seems wrong to base one's interpretation of our fundamental physical theory, QT, on vague notions. And indeed, commentators have criticized the appeal to 'apparatus' and 'pointer position' as vague, and even as question-begging when taken to be part of a solution to the measurement problem (e.g. Stein 1984, pp. 646-647; Bell 1987, pp. 96-97, 124-126; Kent 1990, p. 1754).

Second, one might again object that apparatuses are too few and far between, to secure a definite macrorealm (on a branch): why should the collective relative state for all these pointers being here, there etc. be an eigenstate of the Moon being way over there? In reply, the Everettian might 'bite the bullet' i.e. allow such indefiniteness even within a branch. This would be like the Many Minds approach to defining branches; cf. the comment (Indefinite) at the end of Section 6. (It would also be rather like the way that the Wigner-Stapp version of the (Dynamics) and (DefApp) strategy envisages indefinite macrorealms; cf. Section 5.)

I turn to the third approach. I shall first describe what I take to be the simplest version, and how it can claim to avoid these two objections. That will lead into the further idea that this approach could appeal to anthropocentric notions in its account of preferred systems, and the preferred quantity on them.

The simplest version of this approach takes as the preferred subsystems, all macrosystems in Section 4's sense; i.e. systems that decohere in interaction with their environment. 'Macrosystem' is vague, since 'decoherence kind of interaction' is vague. But it certainly includes all familiar macroscopic objects; and it includes tiny dust particles as well as grosser objects like apparatus pointers, tables, cats and people. As to the preferred quantity on each macrosystem, the approach takes, roughly speaking, the quantity given by the components of the macrosystem's improper mixture after the decoherence interaction. That is, the quantity is to have as its eigenspaces the spaces in terms of which the improper mixture (the macrosystem's density matrix) is very nearly block-diagonalized as a result of the decoherence interaction. This orthogonal family of spaces is often called 'the decoherence basis', 'basis' now being understood more loosely so as to allow multi-dimensional subspaces as well as rays.

This specification of the decoherence basis is vague. Indeed, it is vague in three related ways: as regards the definition of 'decoherence kind of interaction'; as regards when this interaction, which of course never 'turns off', has 'done its stuff'---i.e. when are the off-block-diagonal terms small enough?; and as regards which of the many different quantities that nearly block-diagonalize the density matrix to choose. But vagueness apart, the quantity thus specified is typically 'very close to' a position-like quantity, such as the position of the macrosystem's centre of mass.

So the Everettian now proposes that a branch is specified by taking for every macrosystem, a component of the improper mixture that is its state, together with the relative state (defined by $\Psi$, the state of the universe) of the rest of the universe. So the proposal is similar to the first Many Worlds approach's invocation of N apparatuses, in that an N-tuple of eigenspaces, one from each subsystem's Hilbert space, together with their collective relative state, defines a branch. The difference is that N is now much greater! For the subsystems are now all macrosystems, not just all apparatuses.

As we saw for the first Many Worlds approach, there are various objections, some of which apply to any Everettian strategy, irrespective of the choice of preferred system and quantity (and some of which I will discuss later). And some more specific objections might well be answered (or are artefacts of my exposition!). For example, the approach might well accommodate point (4) above by specifying what a macrosystem is, in a state-dependent way. But what about the two objections against the first approach which I spelt out above?

The first, and obvious, comment concerns the second objection. Namely, the preferred subsystems being all macrosystems, not just all apparatuses ('N being much greater'), means that there is no risk of the collective relative state representing an indefinite macrorealm. A similar comment applies to the first objection, about vagueness. Namely, the amazing efficiency and speed of decoherence mean that the preferred quantity is far more exactly specified than on the traditional Everettian approach, with its mere assumption of a notion of pointer-position: the decoherence basis is very exact by everyday standards. And in reply to this first objection, the Everettian might also urge the considerations of Section 7.1: that for a vague explanandum, one should accept a vague explanans.

This last point returns us to the idea of anthropocentrism. As I mentioned, some proponents of this third approach--- Saunders and Wallace among them---appeal to anthropocentric notions in specifying preferred systems, and the preferred quantity on them. So my simple version of the third approach ---'decoherence applied to all macrosystems'---is set aside. But the details of this appeal of course vary from one author to another, in ways I cannot here survey. So before giving some more details of Saunders and Wallace, it must suffice to make two remarks.

(1): Any such appeal will of course lessen the gap between this third approach and the 'Many Minds' approach. (2): Perhaps the best-known example of such an appeal is in the work of Gell-Mann and Hartle. They adopt the consistent histories approach: which I have eschewed, not least because it builds in time from the beginning---while I think it makes for a clearer exposition to postpone issues about time (till my Section 8). So I cannot here explain their definition of 'consistent history space' (or its special case, 'quasiclassical domain'); which are, roughly speaking, their analogues of 'branch' (or 'world' or 'macrorealm'). Suffice it to say that they relate decoherence to the production of stable records, and thereby to organisms or other macrosystems that exploit the existence of such records so as to survive in their environment (i.e. systems which they call IGUSes---information gathering and utilizing systems); (e.g. Gell-Mann and Hartle 1990).

### 7.3 Saunders and Wallace

At last, I turn fully to 'my' authors, Saunders and Wallace. I shall first briefly discuss their appeal to anthropocentric notions. Then I shall spend most of this subsection developing their arguments against making a once-for-all precise definition of 'branch' or similar words, i.e. of preferred subsystem and/or preferred quantity.

Concerning anthropocentrism: as I mentioned, Saunders casts his views in terms of consistent histories. Indeed, on the present topic---the definition of 'branch' etc., or better, their consistent histories analogues---his views are close to Gell-Mann's & Hartle's (cf. 1995 pp. 240-244, 251-253; 1996a, pp. 128-132). Therein lies the explanation, at least in part, of his resistance to talk of 'branches' and 'worlds'. (He also casts his views in terms of the local algebras approach to quantum theory, which brings to the fore the idea of spatiotemporally bounded events, together with the correlations between them: topics which I will return to briefly, in Section 8.2.)

Wallace, on the other hand, does not adopt the consistent histories approach; (in general, his views are closer to Zurek's and Deutsch's than are Saunders'). Besides, Wallace does accept talk of 'branches' and 'worlds'. In fact, as I mentioned in Section 6 (under (Branch)), he proposes a terminology.

(i) He proposes to use 'branch' for the sort of approximate anthropocentric notion, that he and Saunders agree is needed for securing definite appearances (DefApp). This notion is to be rooted in decoherence theory; (and, he might add, in sympathy with Saunders: perhaps also in the theory of consistent histories). He also requires that branches, so understood, should be temporally extended i.e. not defined only at an instant; and besides, they should evolve stably (in some sense of 'stably'). But exactly how depends branches relate to decoherence, consistency and stability is complicated, and controversial---not least because the physics of decoherence is still being explored.

(ii) Wallace agrees that it is perfectly legitimate to resolve the universal quantum state $\Psi$ into any basis (and to factorize the universe's Hilbert space arbitrarily). He emphasises that although $\Psi$ is specified (and so the universe completely described) by any such resolution, almost all such resolutions are useless. That is, they do not make any empirical facts, whether microscopic or macroscopic, clear or even comprehensible: physicists would call such resolutions 'unphysical' or 'a bad choice of variables'. (Setting aside Saunders' preference for the consistent histories approach, rather than resolutions of an instantaneous state Ψ, he would no doubt agree with these points.) But since all such resolutions of Ψ are perfectly legitimate (though usually useless), it is convenient to have a word for the corresponding 'aspects of the quantum universe (multiverse)'. Wallace proposes to use 'world'. So in this terminology, worlds, unlike the branches, are: precise, defined arbitrarily, and in general defined only at an instant.

So much for Saunders' and Wallace's "main picture". Obviously, one wants to press for more: even if branches (in Wallace's terminology) cannot or even should not be exactly defined, one naturally wants more details about how they secure definite appearances. Equally obviously, these details lie in three directions: the present-day understanding of the physics of decoherence etc., as presented by these and other authors; the future physics of decoherence; and these authors' general or conceptual doctrines about how the branches secure definite appearances.

Clearly, I can only pursue the third direction. More specifically, the rest of this paper will report some central claims and arguments made by both Saunders and Wallace; (again, I postpone their claims about time till Section 8). The overall 'message' of these claims and arguments will be that: (a) we probably will never get a once-for-all precise definition of 'branch' or similar words, i.e. precise definitions of preferred subsystem and/or preferred quantity that are once-for-all best for securing definite appearances; but (b) we do not need such definitions! To report these claims and arguments, I will mainly use 'branch'. That is to say, I will assume that as in (ii) above, the notion of 'world' is taken as precise, arbitrary and in general instantaneous; so that the debate is about whether 'branch' can or should be given a definition as a precise, but also non-arbitrary and temporally extended, notion.

The first claim is that one should *not* take position as the precise preferred quantity. At first sight, position is the obvious candidate, since the measurement problem as originally set out was in effect the threat that macroscopic objects would be in delocalised states. But there are two kinds of reason against it: various technical reasons, and one conceptual one. As to the technical reasons, I think the lack of exact eigenstates, which arises even in elementary quantum theory, is not so important. More important is the fact that the concept of position in relativistic quantum theories, including quantum field theory, is problematic. (Fleming & Butterfield (2000) is a review of the issues, both for relativistic quantum particles and for quantum field theory, in flat spacetime; in more general spacetimes, the problems are worse.)

Nor can one just reply to these difficulties by saying that for quantum field theory, the preferred basis is to be the basis of definite field configurations (which is indeed a natural analogue of elementary quantum theory's notion of particle position). For this basis is also unsuitable for describing macroscopic objects: because the particles of which they are composed are really certain kinds of superposition of these basis states. (Recall the remark in point (4) of Section 7.2: the very existence of the particles that many discussions take as their basic systems is a state-dependent matter.)

The conceptual reason against choosing position relates to what kind of object is to be an 'inhabitant' in a branch. In short, choosing position prevents any object of our familiar ontology, such as dust-particles, apparatus pointers, and people, from being such an inhabitant---it suggests rather that any such object is a collective feature of many branches. In more detail: if we define branches in terms of position---i.e. the position of each and every microscopic constituent, not just a coarse-grained notion like the position of the centre of mass---then the state of any familiar object would be resolved into simultaneous position eigenstates of all its constituents. But the chemical bonds that hold molecules together involve delocalised electrons, so that in any such position eigenstate, the object's electrons are in a superposition of binding and not binding pairs of atoms. So the eigenstate would represent an unstable 'atomic soup' rather than a state of the familiar object: and conversely, any state of the familiar object depends on interference between many different branches. In view of this, it seems wrong to say that nevertheless the object exists in each such branch; instead we should say that the object, like its state, is some kind of collective feature of the branches. (As at the end of the last paragraph, we here meet the idea that the existence of what we normally consider a system, not a state, is in fact state-dependent.)

The second claim is that this conceptual reason against choosing position generalizes. Other apparently natural precise specifications of the preferred quantity, e.g. energy, equally prevent ordinary objects being in a branch, and make them instead collective features of many branches: just because interference between branches thus defined is important to the processes that maintain the object in existence. Presumably, the only way to avoid this problem is to take as the preferred quantity some exact 'decoherence basis', i.e. some member of the vaguely delimited class of quantities selected by decoherence (with the member perhaps also required to satisfy some other constraints). But as I mentioned in Section 4, the physics of decoherence is much more subtle and varied than my summaries suggest: even the vaguely delimited class varies considerably from one kind of physical system (e.g. dust-particle in air or in outer space), or regime for a system (e.g. high or low temperature), to another. This means that it might be unduly restrictive to specify an exact decoherence basis, e.g. in terms of coherent states or localization in phase space---important though these notions are. It also means, on the other hand, that to say just that the preferred quantity is 'some exact decoherence basis' is very vague---in effect a promissory note about future research on decoherence. (Agreed, it is vagueness in the straightforward first sense of Section 7.1, viz. vagueness intended to be temporary.)

Saunders and Wallace conclude from these difficulties that we should 'be liberal' and accept resolutions of the universal quantum state $\Psi$ in an arbitrary basis---or at least an arbitrary basis that is a fine-graining of 'the' decoherence basis. This is their third claim; (setting aside now Saunders' preference for consistent histories, rather than resolutions of an instantaneous state $\Psi$). So in Wallace's terminology of 'worlds': they consider continuously many bases (even if they restrict themselves to bases that fine-grain 'the' decoherence basis), and so are committed to continuously many worlds.

This is certainly a dizzying ontology. After all, each of these continuously many worlds is 'inhabited', i.e. each world is not just a component of a state (where states represent reality) but also has a 'system' (albeit not an ordinary object!) actually in it. (Besides they would still have continuously many worlds even if for each basis they said---as they does not---that only one world is 'inhabited'.) But this dizzying ontology---this dizzying pill for curing us of the measurement problem---is sugared by three other considerations, (A) to (C) below: of which the third is most important, and will lead us to Section 8.

(A) The first consideration is due to Saunders rather than Wallace; (though I believe they agree on the point). It relates to the interpretation of probability, a topic which I foreswore in Section 1, both for reasons of space and because I largely agree with Saunders. So here a very short explanation must suffice. Some Everettians propose that each component of $\Psi$ in a preferred quantity is inhabited by a whole population of systems: for only with a population, they say, can we make sense of various different probabilities for future alternatives. The idea is that given a population, 'more of them' can transit to future alternative A than to B; and this makes sense of A being more probable. This proposal is largely independent of whether one takes there to be just one preferred quantity, as traditionally; or many, as in Saunders and Wallace's 'liberalism'. (I say 'largely', rather than 'entirely', because of a point stressed by Hemmo & Pitowsky (2001): that if one takes there to be more than one preferred quantity, and they do not commute, then to prevent commitment to a violated Bell inequality, one will have to allow a sort of contextuality about how systems are assigned to the various populations.) This proposal is also independent of whether any such preferred quantity is specified in physical terms (Many Worlds) or in mental terms (Many Minds): the claim is made by advocates of both these approaches (e.g. Deutsch (1996), and Lockwood (1996) respectively).

But I (and Saunders, and I believe Wallace) reject this proposal, as a naïve attempt to ground probability in counting measures, or in some version of the principle of indifference. These are both dead horses in the philosophy of probability---which I will not flog; (Saunders 1996: Section 6, and 1998; Butterfield 1996, Section 5). Suffice it here to make two points.

First, rejecting this proposal means we believe the Everettian can maintain there is only one system in each world, so that systems indeed 'split' when the number of worlds increases; cf. Section 8 for more discussion. Second, the proposal does *not* sit very well with the Everettian's main idea that all the worlds are actual, 'inhabited'. For suppose there is no splitting, but only a stochastic process, continually re-partitioning an overall population of persisting systems; then it is hard to see why the Everettian should postulate more than one system, i.e. more than one world being 'inhabited'. (This, I believe, lies behind some authors' suggestions that the Everettian can and should avoid this postulate, and instead be like the pilot-wave interpretation in postulating just one actually possessed value of the preferred quantity; cf. Bell (1987, p. 97, p. 133).)

(B) The second 'sugar on the pill' lies in what I called the conceptual reason against specifying position to be the preferred quantity: that any familiar object is a collective feature of many worlds. Putting this the other way around: the 'system' inhabiting a world is not a familiar object---unless the world is defined by one of the (vaguely delimited!) class of 'decoherence bases'; (i.e. in Wallace's terminology: unless the world is a branch).

(C) The third sugar on the pill is an analogy between worlds and the spacelike slices of a relativistic spacetime, as conceived on the 'tenseless' or 'block universe' conception of time. This analogy builds on that between worlds and times that I gave in (Branch) in Section 6, as the Everettians' answer to the charge of contradiction.

The idea of the analogy is that just as someone who accepts the tenseless conception of time can readily accept instants i.e. spacelike slices of spacetime, as (i) useful or even indispensable for describing phenomena, and yet (ii) not any substantive ontological commitment additional to the commitment to spacetime; so also an Everettian can readily accept worlds as (i) useful or even indispensable, and yet (ii) not a substantive commitment additional to the commitment to actuality's being described by the universal state.

This turns out to be a rich and multi-faceted analogy. It is developed in detail by Saunders (1995, Sections 1, 7; 1996) and Wallace (2001); as Saunders notes, it goes back at least to Geroch (1984). But I must confine myself to filling out the main idea, and mentioning a few facets.

Recall that a classical relativistic universe can be described as a collection of instants (spacelike slices) and their temporal relations to each other. There is great arbitrariness in specifying these slices (i.e. in how spacetime is foliated); and any foliation breaks some symmetries of the spacetime. But any foliation, together with its slices' temporal relations, gives a complete description of the universe. And in everyday circumstances, or in treating some specific problem in physics, there is an approximately defined best choice of foliation—though the details are arbitrary, especially as regards how to foliate spacetime far away (in space or in time) from the region whose events one aims to describe.

I emphasise for clarity that the Everettian analogue will concern the quantum universe *at a time*. So the Everettian says: similarly a quantum universe *at a time* can be described as a collection of worlds and their amplitudes $c_i$. There is great arbitrariness in specifying these worlds, i.e. in specifying a basis, or more generally quantity (complete orthogonal family of subspaces); and any specification breaks some symmetries of the quantum universe at the time. But any choice of basis (i.e. of a maximal quantity) together with its elements' amplitudes, gives a complete description of the quantum universe at the time. And in treating some specific problem in quantum physics (and in everyday circumstances, to the extent that quantum theory is needed for them!), there is an approximately defined best choice of basis, or more generally quantity—though the details are arbitrary, especially as regards how to specify the quantity for events other than those one aims to describe.

As the parallel wordings show, this is a telling analogy—suggesting that Everettian worlds should be as innocuous to someone who 'takes quantum theory seriously' as instants are according to the tenseless conception of time. Furthermore, the analogy has other facets. Some of these concern the Everettians' treatment of time, so I postpone them to Section 8. But I should mention here two examples.

(1): Just as the tenseless conception of time must somehow reconcile its account with the intuitive but problematic 'flow of time'; so also the Everettian must somehow reconcile her account of a deterministically evolving $\Psi$ with the intuitive but problematic notion of probability. But in this paper I have foresworn the interpretation of probability: as I said above and in Section 1, I largely agree with Saunders' views.

(2): This facet concerns the treatment of time---but by classical physics and metaphysics, not by Everettians. They provide an analogue of the idea above, that any familiar object is a collective feature of many worlds; (which militated against specifying position as the preferred quantity, and suggested that the inhabitants of worlds were 'unfamiliar systems'). The analogue is clear. The states of objects, as we conceive them both in everyday life and in science, are usually not instantaneous. We say that the gun fires, the vial breaks, the cat falls: each requires the object to persist for some (vaguely defined!) period of time. Similarly, for mental states: to see yellow, an organism must be conscious for, say, a fifth of a second. Agreed, science, and especially physics, has fashioned concepts of instantaneous state, which have been supremely successful. (Indeed, so successful that nowadays, physicalism is often combined with my point about mental states: so people say that mental states do not supervene on instantaneous physical states). But for all their success, or even fundamentality, instantaneous states are a minority; most of what we call 'states of objects' make no sense as holding only for an instant.

## 8. The Treatment of Time

The Everettian's treatment of time involves various issues, both philosophical and technical. I must confine myself to a brief discussion; (which complements my (1995 Sections 6, 7; 1996 Section 4)). In Section 8.1, I shall state another facet of Section 7.3's closing analogy; this will lead to related philosophical issues about identity through time. I will conclude in Section 8.2 with a few remarks about dynamics.

### 8.1 Issues about Identity through Time

To state the new facet of section 7.3's analogy, I again start on the 'classical spacetime side'. We have no exact criterion of identity over time for familiar objects, like dust particles, tables, cats and people; but we can nevertheless make do both in everyday and scientific life (everywhere but philosophy's puzzle-cases of teleportation, fission, brain-transplant etc.) with vague and approximate criteria of identity. So also, says the analogy, the Everettian can admit to having no exact criterion of identity over time for the systems that are inhabitants of worlds, the vast majority of which, as we have seen, are not familiar objects; but the Everettian can nevertheless make do with vague and approximate criteria of identity.

To explain this, recall first that ever since Section 6, the Everettian's worlds have been defined only at an instant: although the word 'world' (and equally 'branch', 'system' and 'inhabitant') suggests persistence over time, the discussion hitherto has not needed any Everettian account about what such persistence is to involve: in philosophers' jargon, what worlds' criteria of identity are. The present suggestion, endorsed equally by Saunders and Wallace, is that the Everettian need not give a uniform account, the same for all worlds (or branches etc.). Nor need she give a precise but non-uniform account: i.e. an account that first divides worlds etc. into different kinds, and then gives different exact accounts for different kinds.

Indeed, for Saunders and Wallace, she *should* not do so. For as we have seen, they allow arbitrary resolutions of $\Psi$ to define worlds; and almost all such resolutions are 'unphysical'. In particular, they will be useful for describing macroscopic facts only if they are a 'decoherence basis', i.e. some member of the vaguely delimited class of quantities selected by decoherence; (maybe they could also be moderate fine-grainings or coarse-grainings of a decoherence basis). So of course we should not expect a useful notion of persistence to be definable for arbitrary worlds.

On the other hand, a few of these resolutions, viz. the decoherence bases (and maybe also moderate fine-grainings or coarse-grainings of them), describe familiar macroscopic objects, from dust particles to people. So we *should* expect to be able to define persistence for the worlds given by these resolutions. Or rather, we should be able to do this vaguely and approximately, just as we do for dust particles to people. In short: the suggestion is that persistence makes no sense for arbitrary worlds, but makes vague and approximate sense for macroscopic objects (in the way familiar from philosophical discussions of identity over time).

All this I find plausible. But the topic of worlds' criteria of identity over time also raises other issues which are not part of Section 7.3's analogy: issues to do with the 'splitting' of worlds over time. Although the Everettian has defined worlds only relative to an instant, she of course must accept different numbers of worlds at different times: this occurs whenever an eigenstate unitarily evolves into a superposition. For example, in Section 3's toy-model of measurement, there was one world before the measurement interaction, and two after. Hence the widespread use of the word 'branch', based on the image of a tree-trunk splitting into branches. But how exactly should we understand such splitting?

The first point to make is a matter of ground-clearing. I agree that Everettians who hold that each component of $\Psi$ in a preferred basis is inhabited by a whole population of systems can avoid the number of worlds changing in time. For they can say that different members of such a population evolve to different later components: they simply 'go different ways' during a so-called 'splitting'. But in Section 7.3, I already joined Saunders and Wallace in rejecting the idea of each component being inhabited by a whole population, for it seems motivated only by a bad reason---the attempt to ground probabilities in counting, or in some version of the principle of indifference. So my question is: assuming that at any one time there is only one 'inhabitant' in each world, how should we understand splitting?

Again, Saunders has addressed this question: partly by comparing it with the question, much studied in philosophical discussions of identity over time (especially personal identity), of how to understand cases of fission of ordinary objects (including people); and partly by connecting it to his discussion of probability (1998, Sections 4-6). Again, I find his account plausible, but cannot enter into details. Yet there is one aspect of the question which is independent of the above discussion about allowing vague criteria of identity, and yet so central to the topic of identity over time (and so closely related to the analogy between worlds and times, even in its basic form from (Branch) of Section 6, rather than from Section 7.3)---that I must address it.

This aspect turns on the fact that a well-known philosophical dispute about the identity of objects over time has a clear analogue for the Everettian's worlds. Recall that philosophers treat change of properties over time in two rival ways, depending on whether they conceive objects that persist over time as (a) strictly self-identical across time, or as (b) having stages (also known as temporal parts). It will be convenient to adopt a recent jargon used in the dispute over temporal parts (cf. Lewis 1986, p. 202). Namely, let us say that (i) an object that persists by having temporal parts at various times 'perdures'; and (ii) an object without temporal parts, i.e. that is wholly present at different times, 'endures';

and finally, (iii) the neutral word, covering perdurance or endurance, is 'persist'. I will use 'transtemporal identity' and 'persistence' as synonyms.

These distinctions suggest similar ones applying to worlds. Should we say that: (a) one object, say the pointer, has contrary properties, say two non-overlapping positions, relative to distinct worlds or branches? Or should we say that: (b) there is more than one object, which are very similar to one another (hence often called 'copies') but for having different positions?

Most Everettians adopt the second option, (b); and besides, so do Saunders and Wallace. I agree that there are good reasons to do so: (reasons which hold good, even if you resist the temporal analogue, i.e. resist temporal parts). But it is important to point out the first option, (a); (cf. Tappenden (2000), who suggests that we can think of an object relative to a world as a new kind of part of the object, which he calls a 'superslice').

For we should beware of two *bad* reasons for adopting (b). First: for many Everettians, (b) is simply a corollary of their view (which Saunders, Wallace and I reject) that each component of $\Psi$ in a preferred basis is inhabited by a whole population of systems. But note that the converse fails: i.e., (b) does not require their view, so that Saunders and Wallace can combine advocating both (b) and genuine 'splitting'. In particular, there is no conflict between this combination and conservation laws, like the conservation of mass or of charge: we should *not* say that when a pointer 'splits' during a measurement-process to give two pointers in non-overlapping positions, the mass is doubled. For after all, we do not say in the temporal analogue that the advocate of perdurance, who believes that a pointer of mass one kilogram has two temporal parts, is thereby committed to the existence of a two kilogram mass.

Secondly, most discussions ignore the point made in Section 7.3, that on any exact definitions of world, any ordinary objects are collective features of many worlds. (Indeed for the sake of clarity, I just did so myself, when presenting the choice between (a) and (b) as a matter of whether there are one or two pointers.) Once we notice this point, (a) is more plausible: if an object is 'extraordinary', why not let it have contrary properties relative to different worlds (or even corresponding parts as Tappenden advocates)?

So what it is the good reason for (b)? I think it is that the contrary properties involved in two alternative outcomes of a measurement process might well be so much more disparate than two non-overlapping positions, that it seems wrong to say some single object is involved. I have in mind not just such pairs of properties as being alive, and being dead from poisoning, in the case of Schrödinger's proverbial cat; here one might well treat the body (at least for a while) as the same object in the two outcomes, alive and dead. But there are more extreme cases. Imagine that one of the two outcomes of a measurement vaporizes the pointer: what is the common object then?

You might suggest that we should identify an object with its microscopic constituents; so that in the case of vaporizations etc. we should trace these in space and time. But this suggestion faces objections both from philosophy and from physics. As I see it, the main philosophical objection is that the suggestion implies wrongly that the pointer survives being vaporized. The main physical objection arises from the point made in Section 7.3, that quantum field theory treats particles as excitations of fields: so when we try to make 'microscopic constituents' precise, we must face the fact that the very existence of the particles we take as constituting objects, in fact depends on the quantum state (of the field-system)---and this state can vary between Everettian worlds in such a way that a particle can exist in one but not the other of two worlds after a 'split'.

### 8.2 Issues about Dynamics

Finally I turn to a more technical topic: what the Everettian says about dynamics, i.e. the laws of evolution for quantum states. I will just make brief remarks, first about determinism and then about relativistic covariance. (For these issues, the metaphysical concerns of Section 7.3 and 8.1, such as the inhabitation of branches and the nature of persistence through time, are by and large irrelevant.)

First, we need to distinguish two main areas of discussion: the dynamics of the universe as a whole, and the dynamics of subsystems of the universe. The former is simpler---at this paper's high level of generality. That is: of course the detailed physics of the universe as a whole is impossibly complex, and will surely remain forever beyond our ken. But the fact that the universe is an isolated system (strictly speaking, the only one) means that according to orthodox quantum theory---and the Everettian---it (and it alone) obeys the deterministic unitary evolution of the Schrödinger equation. On the other hand, subsystems of the universe are not isolated (they are 'open'); i.e. they are influenced by their environment, so that their evolution is non-unitary. In general, the theoretical description of their evolution is much more complex than that of isolated systems. So, as is often remarked, Schrödinger and the other pioneers of quantum theory were lucky that nature supplied some subsystems of the universe (viz. atoms) that were both sufficiently isolated that the Schrödinger equation is an excellent approximation, and sufficiently simple that at least some aspects of their behaviour could be understood (at least by a pioneer of genius!) (Similarly of course for Newton and his contemporaries: they were lucky that nature supplied sufficiently isolated and sufficiently simple subsystems of the universe---the ball on the inclined plane, the planets.)

Thus we return to one of Section 1's opening themes: determinism vs. indeterminism. For the Everettian's idea will of course be that her postulated deterministic evolution of the whole universe is compatible: with (i) indeterministic evolution of generic subsystems---which is only to be expected since the subsystems are open; and also with (ii) almost deterministic evolution of subsystems that are almost isolated, e.g. a molecule in a gas. How to recover from an over-arching unitary evolution the right type of evolutions of kinds (i) and (ii) is of course an ongoing research area in physics (in which decoherence plays important roles).

I turn to relativistic covariance. It is of course intended as a principal merit of Everettian interpretations that they obey relativistic covariance, fundamentally rather than just at a phenomenological level; (unlike the pilot-wave interpretation, and maybe also the modal interpretation and theories with a physical process of wave-packet collapse; cf. Section 5). But again, it is important to distinguish the dynamics of the universe as a whole, from the dynamics of its subsystems. So far as I know, Everettian discussions almost always focus on the relativistic covariance of the former; and as I shall spell out in a moment, I see no problems of principle lurking there. But since of course the relativistically covariant theories we in fact confirm are theories of subsystems of the universe (and very simple and tiny ones, to boot!), the Everettian also owes us an account of how her postulated relativistically covariant evolution of the universe yields relativistically covariant evolution for appropriate subsystems. So here is a lacuna, albeit one that may be easy to fill. But let me end more positively, by outlining how the Everettian's dynamics of the universe as a whole can be fundamentally relativistically covariant---and thereby returning to Saunders.

Relativistic covariance is obscured by the traditional way of presenting the Everettian strategy which I also adopted in Sections 6 and 7; viz. using relative states, defined in terms of the instantaneous state $\Psi$. But there is no real problem here, for three reasons; I give them in increasing order of technicality---the first two in effect giving us two different notions of relativistic covariance.

First, a fundamentally relativistic quantum theory can perfectly well use foliations of spacetime, with states $\Psi$ associated with slices; provided it allows the use of arbitrary spacelike foliations, with the $\Psi$ on slices of different foliations being suitably related (in particular by a unitary transformation). Second, there is an approach to relativistic quantum theories which lends itself easily to foliation-independent notions and techniques: viz. the local algebras approach, which associates an algebra of quantities with each bounded spacetime region. And third, there is an approach to all quantum theories (including relativistic ones), viz. the consistent histories approach, in terms of which one can formulate the Everettian strategy, using a generalized and foliation-independent notion of relative state.

Much more can be, and has been, said about these three reasons; in particular about the relations between them. Here I can only note that Saunders combines these last two approaches---local algebras and consistent histories: in short, the projectors that define the histories are taken to be elements of local algebras. In particular, he defines a relation of 'relative definiteness' between histories $H_1$ and $H_2$, which is roughly '$H_1$ has probability 1, given that $H_2$ is realized'; and then he discusses the relation of the consistency condition on histories to properties like the symmetry and transitivity of relative definiteness; (1995, Sections 2, 3; 1996a, Section 3). But I must forego details: for now, this must suffice by way of urging that there are good prospects for formulating the Everettian strategy in relativistic quantum theories, and in particular in the local algebras approach.

*Acknowledgements*:--Many thanks to the editors and conference participants, to Meir Hemmo, Adrian Kent, Paul Tappenden and Lev Vaidman---and especially to Simon Saunders and David Wallace---for discussion and comments on previous versions: if only I could have consistently acted on all suggestions!

**References**

Barrett J., 1999. *The Quantum Mechanics of Minds and Worlds*, Oxford: Oxford University Press.

Bell J., 1987. *Speakable and Unspeakable in Quantum Mechanics*. Cambridge: Cambridge University Press.

Berry, M., 1994. Asymptotics, singularities and the reduction of theories, *Proceedings of 9th International Congress of Logic, Methodology, and Philosophy of Science*, eds. D. Prawitz, B. Skyrms and D. Westerståhl, Dordrecht: Kluwer, pp. 597-607.

Berry, M., this volume: Chaos and the semiclassical limit of Quantum mechanics (is the moon there when somebody looks?) *Proceedings of CTNS-Vatican conference on Quantum Physics and Quantum Field Theory*.

Bohm D. & Hiley B., 1993. *The Undivided Universe: an Ontological Interpretation of QuantumTheory*. London: Routledge.

Bub J., (1997), *Interpreting the Quantum World*, Cambridge: Cambridge University Press.

Butterfield J., 1995. Worlds, Minds and Quanta, *Aristotelian Society Supplementary Volume* **69**, pp. 113-158.

Butterfield J., 1996. Whither the Minds?. *British Journal for the Philosophy of Science* **47** pp. 200-221.

Butterfield J., 1996a. Quantum Curiosities of Psychophysics, in *Human Consciousness and Identity*, ed. J. Cornwell, Oxford: Oxford University Press, pp. 123-159. Pittsburgh Philosophy of Science Archive: http://philsci-archive.pitt.edu/ PITT-PHIL-SCI00000193

Butterfield, J. 2001. The State of Physics; Halfway through the Woods, *Journal of Soft Computing*, **5**, 2001, pp. 129-130.

Butterfield, J., & Isham, C., 1999. The Emergence of Time in Quantum Gravity, in *The Arguments of Time*, ed. J. Butterfield, London: British Academy and Oxford University Press. eprint: gr-qc/9901024

Clarke, C. this volume. Title? *Proceedings of CTNS-Vatican conference on Quantum Physics and Quantum Field Theory*.

Cushing J., 1994. *Quantum Mechanics: Historical Contingency and the Copenhagen Hegemony*. Chicago: University of Chicago Press.

Cushing, J. this volume, Title? *Proceedings of CTNS-Vatican conference on Quantum Physics and Quantum Field Theory*.

Dieks D. & Vermaas P., eds. (1998), *The Modal Interpretation of Quantum Mechanics*, Dordrecht: Kluwer.

Deutsch, D. 1996. *The Fabric of Reality*, London; Penguin.

Dowker F., & Kent A., 1996. On the Consistent Histories Approach to Quantum Mechanics, *Journal of Statistical Physics* **82** pp. 1575-1646.

Everett H., 1957. 'Relative State' Formulation of Quantum Mechanics, *Reviews of Modern Physics* **29**, pp. 454-462.

Fleming, G. & Butterfield, J. 2000. Strange Positions, in *From Physics to Philosophy*, ed. J. Butterfield and C.Pagonis, Cambridge: Cambridge University Press, pp. 108-165.

Gell-Mann M., & Hartle J., 1990. Quantum Mechanics in the light of Quantum Cosmology, in *Complexity, Entropy and the Phyiscs of Information*, ed. W. Zurek, Reading: Addison-Wesley, pp. 425-459.

Geroch R., 1984. The Everett Interpretation, *Nous* **18**, pp. 617-633.

Ghirardi G., Grassi R. & Benatti F., 1995. Describing the Macroscopic World: Closing the Circle in the Dynamical Reduction Program, *Foundations of Physics* **25**, pp. 5-40.

Giulini D., et al. 1996. *Decoherence and the Emergence of a Classical World from Quantum Theory*, Berlin: Springer.

Hemmo M., & Pitowsky I., 2001. Probability and Nonlocality in Many Minds Interpretations of Quantum Mechanics. In preparation.

Holland P., 1993. *The Quantum Theory of Motion*, Cambridge: Cambridge University Press.

Joos E. & Zeh H.D., 1985. Emergence of Classical Properties through Interaction with the Environment. *Zeitschrift für Physik* B**59**, pp. 223-235.

Kent A., 1990. Against Many Worlds Interpretations, *International Journal of Modern Physics* A**5**, pp. 1745-1762.

Lewis D., 1973. *Counterfactuals*, Oxford: Blackwell.

Lewis D., 1986. *On the Plurality of Worlds*, Oxford: Blackwell.

Lockwood M., 1996. 'Many-Minds' Interpretation of Quantum Mechanics, *British Journal for the Philosophy of Science* **47** pp. 159-188.

Redhead, M. this volume. The Tangled Story of the Nonlocality Issue in Quantum Mechanics, *Proceedings of CTNS-Vatican conference on Quantum Physics and Quantum Field Theory*.

Saunders S., 1995. Time, Quantum Mechanics, and Decoherence, *Synthese* **102**, 235-266.

Saunders S., 1996. Time, Quantum Mechanics, and Tense, *Synthese* **107**, pp. 19-53.

Saunders S., 1996a. Relativism, in *Perspectives on Quantum Reality*, ed. R. Clifton, Dordrecht; Kluwer, pp. 125-142.

Saunders S., 1998. Time, Quantum Mechanics, and Probability, *Synthese* **114**, pp. 373-404

Stapp H., 1993. *Mind, Matter and Quantum Mechanics*, New York; Springer-Verlag.

Stein H., 1984. The Everett Interpretation of Quantum Mechanics: Many Worlds or None?, *Nous* **18**, pp. 635-652.

Tappenden P., 2000. Identity and Probability in Everett's Multiverse, *British Journal for the Philosophy of Science* **51** pp. 99-114.

Vaidman L., 1998. On Schizophrenic Experiences of the Neutron, Or Why we Should Believe in the Many-Worlds Interpretation of Quantum Theory, *International Studies in Philosophy of Science* **12**, pp. 245-261; eprint quant-ph/9609006.

Vaidman L., 2000. The Many-Worlds Interpretation of Quantum Mechanics, *The Stanford Encyclopedia of Philosophy;* only on web at http://plato.stanford.edu/.

Vermaas P., 2000, *A Philosopher's Understanding of Quantum Mechanics: Possibilities and Impossibilities of a Modal Interpretation*, Cambridge: Cambridge University Press.

Wallace D., 2001. Worlds in the Everett Interpretation, forthcoming in *Studies in History and Philosophy of Modern Physics.* eprint: quant-ph/0103092.

Wallace D., 2001a. Thoughts on Quantum Mechanics and vice versa: Fitting the Observer in to the Interpretation of Quantum Theory. In preparation.

Wigner E., 1962. Remarks on the Mind-Body Problem, in Good I.J. (ed.), *The Scientist Speculates*, London: Heinemann.